\documentclass[sn-mathphys,Numbered]{sn-jnl}% Math and Physical Sciences Reference Style
\usepackage{graphicx}%
\usepackage{multirow}%
\usepackage{amsmath,amssymb,amsfonts}%
\usepackage{amsthm}%
\usepackage{mathrsfs}%
\usepackage[title]{appendix}%
\usepackage{xcolor}%
\usepackage{textcomp}%
\usepackage{manyfoot}%
\usepackage{booktabs}%
\usepackage{algorithm}%
\usepackage{algorithmicx}%
\usepackage{algpseudocode}%
\usepackage{listings}%
\usepackage{placeins} % with \FloatBarrier, it keeps figures inside a subsection
\theoremstyle{thmstyleone}%
\theoremstyle{thmstyletwo}%

\theoremstyle{thmstylethree}%

\begin{document}

\title[Article Title]{Thermal Earth Model for the Conterminous United States Using an Interpolative Physics-Informed Graph Neural Network (InterPIGNN)}


\author*[1]{\fnm{Mohammad J.} \sur{Aljubran}}\email{aljubrmj@stanford.edu}

\author[1]{\fnm{Roland N.} \sur{Horne}}\email{horne@stanford.edu}

\affil*[1]{\orgdiv{Energy Science \& Engineering}, \orgname{Stanford University}, \orgaddress{\street{Building 367}, \city{Stanford}, \postcode{94305}, \state{California}, \country{United States}}}


\abstract{This study presents a data-driven spatial interpolation algorithm based on physics-informed graph neural networks used to develop national temperature-at-depth maps for the conterminous United States. The model was trained to approximately satisfy the three-dimensional heat conduction law by simultaneously predicting subsurface temperature, surface heat flow, and rock thermal conductivity. In addition to bottomhole temperature measurements, we incorporated other physical quantities as model inputs, such as depth, geographic coordinates, elevation, sediment thickness, magnetic anomaly, gravity anomaly, gamma-ray flux of radioactive elements, seismicity, and electric conductivity. We constructed surface heat flow, and temperature and thermal conductivity predictions for depths of $0$-$7 \ km$ at an interval of $1 \ km$ with spatial resolution of $18 \ km^2$ per grid cell. Our model showed superior temperature, surface heat flow and thermal conductivity mean absolute errors of $4.8^\circ C$, $5.817 \ mW/m^2$ and $0.022 \ W/(C \cdot m)$, respectively. The predictions were visualized in two-dimensional spatial maps across the modeled depths. This thorough modeling of the Earth's thermal processes is crucial to understanding subsurface phenomena and exploiting natural underground resources. The thermal Earth model is made available as feature layers on ArcGIS at \href{https://arcg.is/nLzzT0}{https://arcg.is/nLzzT0}.}

\keywords{Temperature-at-Depth; Heat Flow; Rock Thermal Conductivity; InterPIGNN; Physics-Informed; Graph Neural Networks}

\maketitle

\section{Introduction}\label{sec1}
Understanding temperature variations across depths and geographies holds significance to many domains. Temperature is fundamental to modeling heat distribution, thermal anomalies, groundwater flow and management, volcanic eruptions, stresses and earthquakes, hydrocarbon maturity, carbon sequestration, subsurface energy storage, amongst others \cite{kukkonen1994subsurface, chen2016bedrock, vilarrasa2017thermal, head2003biological}. In particular, accurate modeling of temperatures at significant depths within the Earth's crust is central to the exploration and development of geothermal energy. A conventional geothermal resource requires three ingredients: (1) heat source, (2) fluid for heat transport, and (3) permeable/fractured rock for fluid flow. Other unconventional geothermal systems can also be viable based on site-specific circumstances \cite{dipippo1991geothermal, mock1997geothermal}. Specifically, enhanced geothermal systems (EGS) only require sufficiently high-temperature resources as heat transport and fluid flow can be achieved through water injection and rock stimulation, respectively \cite{kana2015review}. Many recent efforts were focused on assessing the EGS resource potential across the United States \cite{blackwell2006assessment, augustine2016update, augustine2019geovision, tester2006future} using either regional or national temperature-at-depth maps.

Regional temperature-at-depth maps were developed for various locations. One study by researchers at Southern Methodist University (SMU) Geothermal Laboratory modeled heat flow of the Appalachian Basin using two-dimensional finite difference heat conduction, basin cross sections, equilibrium temperature, and oil and gas bottom-hole temperature (BHT) data to quantify heat flow at the surface and at the base of the sedimentary basin. Other efforts by researchers at Cornell University used heat flow estimates based on BHT measurements to estimate temperature-at-depth at the Appalachian Basin using a one-dimensional heat conduction model \cite{smith2019appalachian} coupled with Monte Carlo analysis of various physical quantities for uncertainty quantification. SMU researchers also developed a regional temperature-at-depth model for the Cascades using a three-dimensional finite difference heat conduction model \cite{frone2015shallow}. Other thermal models were focused on the Great Basin and led by multiple institutes, e.g., Nevada Bureau of Mines and Geology, US Geological Survey (USGS), and INGENIOUS project \cite{coolbaugh2005geothermal, deangelo2023new, ayling2022ingenious}. Others estimated temperature-at-depth and subsurface thermal properties at the Snake River Plain using field measurements alongside a conductive and radiogenic heat transfer and production model. \cite{batir2020shallow, nielson2012geothermal}. Whereas such regional models are useful, they are spatially limited and vary in input quantities, assumptions, BHT accuracy, spatial and depth-wise resolutions, inconsistency if integrated, amongst other aspects. This prompts the need for a national temperature-at-depth models to guide conventional and non-conventional geothermal exploration and development.

National temperature-at-depth maps for the conterminous US were developed by the SMU Geothermal Laboratory as part of a 2006 study on \textit{The Future of Geothermal Energy}, mainly focused on assessing the potential of EGS across the US \cite{blackwell2011temperature, tester2006future, blackwell2006assessment}. However, the SMU maps have several limitations:
\begin{itemize}
    \item They were developed at a gridding interval of 5 minutes (or, $0.08333^{\circ}$) of spatial resolution, which translates to an average spacing of about 8 km representing an area of about $64 \ km^2$ per grid cell. A typical 250 MWe EGS plant might require about $10-20$ $km^2$ of reservoir planar area to accommodate the thermal resource needed, assuming that heat removal occurs in a 0.5 km-thick region of hot rock at depth \cite{tester2006future}. With such a large areal extent, a $64 \ km^2$ grid cell is likely to filter out many local heat anomalies. These maps also ignore EGS resources shallower than 3 km entirely, and only model depth intervals between 3-10 km \cite{augustine2016update}.
    \item The analytical approach used in developing these temperature-at-depth maps requires knowledge of heat flow across the conterminous US. Note that all hydrothermal system-influenced data (very high values, i.e., generally greater than $120 \ mW/m^2$ were conveniently excluded in this process to avoid the difficulty involved in analytically modeling regions with hot fluid upflow without overestimating temperatures at neighbouring sites. This inconveniently leads to smoothing many local heat anomalies which developers would rather like to capture for economic EGS development.
    \item In developing such heat flow maps, a minimum curvature algorithm was used. However, such interpolation algorithms (1) are less effective with irregularly spaced data points, and (2) tend to generate smooth fits that dismiss potential heat anomalies in undersampled regions.
    \item SMU adopted a two-layer model where the sediment and basement contributions were captured separately. This requires knowledge of either unavailable/assumed or interpolated spatial properties across layers, such as rock thermal conductivity, measured sediment heat flow, basement heat flow, radioactive depth variable, and radioactive heat generation. 
\end{itemize}

Interested in estimating shallow, low-temperature geothermal resources across the conterminous US, researchers at the National Renewable Energy Laboratory developed a temperature-at-depth model to capture depths shallower than $3 \ km$ \cite{mullane2016estimate}. Approximately 300,000 BHT measurements were used in that study. To reduce the nugget effect associated with the variogram of the kriging algorithm used in that study, the BHT observations were calibrated to the nearest $500 \ m$ depth interval using geothermal gradients obtained from the $3.5$-$km$ temperatures originally predicted by the SMU temperature-at-depth maps. Using a grid of $16 \ km^2$, BHT measurements were sampled down to approximately 71,000 records per grid cell using the median statistic. Then, they fit a linear regression model to predict BHT using (1) depth, and (2) the $3.5$-$km$ temperatures originally predicted by the SMU temperature-at-depth maps. Next, they applied ordinary kriging to the linear regression residual which were used to improve local fit. In addition to the aforementioned caveats associated with the SMU temperature-at-depth maps, these shallower maps lack spatial and depth-wise resolution due to downsampling real BHT measurements. Our investigation shows that subtracting the $3$-$km$ temperature-at-depth map developed in this study from the $3.5$-$km$ temperature-at-depth map developed by SMU results in an average difference of over $-50^{\circ} \ C$.

This study presents an \textbf{I}nterpolative \textbf{P}hysics-\textbf{I}nformed \textbf{G}raph \textbf{N}eural \textbf{N}etwork (\textbf{InterPIGNN}) used to develop national temperature-at-depth maps for the conterminous US. In addition to bottomhole temperature measurements, we incorporated other physical quantities as model inputs, such as depth, geographic coordinates, elevation, sediment thickness, magnetic anomaly, gravity anomaly, gamma-ray flux of radioactive elements, seismicity, and electric conductivity. Data curation, preprocessing, and normalization are significant steps in this process. We trained, validated, and tested our model on unseen BHT measurements to evaluate generalizability. We then used it to predict temperature-at-depth for depths of $0$-$7 \ km$ at an interval of $1 \ km$ with spatial resolution of $18 \ km^2$ per grid cell that is sufficient for typical EGS projects.

\section{Data Collection}\label{sec2}
Various physical quantities and data sources are required toward an accurate estimate of temperature-at-depth across the conterminous US. Measurements span various geographic locations (Easting/Northing) and/or depth. We collected measurements for three thermal quantities: BHT, heat flow, and rock thermal conductivity. Additionally, we allocated measurements with sufficiently high spatial resolution of other physical quantities which can be explicitly or implicitly correlated with temperature-at-depth, i.e., average surface temperature, elevation, sediment thickness, magnetic anomaly, gravity anomaly, Uranium radiation, Thorium radiation, Potassium radiation, seismicity, and electric conductivity. Note that these measurements have various spatial resolutions, hence we downsampled/upsampled them to populate the continental United States with the target resolution of $18 \ km^2$ using inverse distance weighing \cite{lu2008adaptive}. Note that all visualizations reported in this article were generated by the authors accordingly. In the following, we briefly describe the source and distribution associated with these quantities.

\subsection{Thermal Quantities}\label{sub2.1}
BHT measurements are acquired through direct wellbore temperature measurements at one or more depths. The collected temperature measurements could be associated with geothermal, oil \& gas, groundwater, observation, and monitoring wells. After exhaustive search and curation of BHT measurements, we identified nine different raw and aggregated data sources which we integrated into a single database. These sources are SMU, Association of American State Geologists (ASSG), USGS, Utah Geological Survey (UGS), Colorado Geological Survey (CGS), Maine Geological Survey (MGS), Washington State Department of Natural Resources (WSDNR), Geothermal Information Layer for Oregon (GTILO), and Great Basin Center for Geothermal Energy (GBCGE).

\begin{figure}[h!]%
\centering
\includegraphics[width=1.0\textwidth]{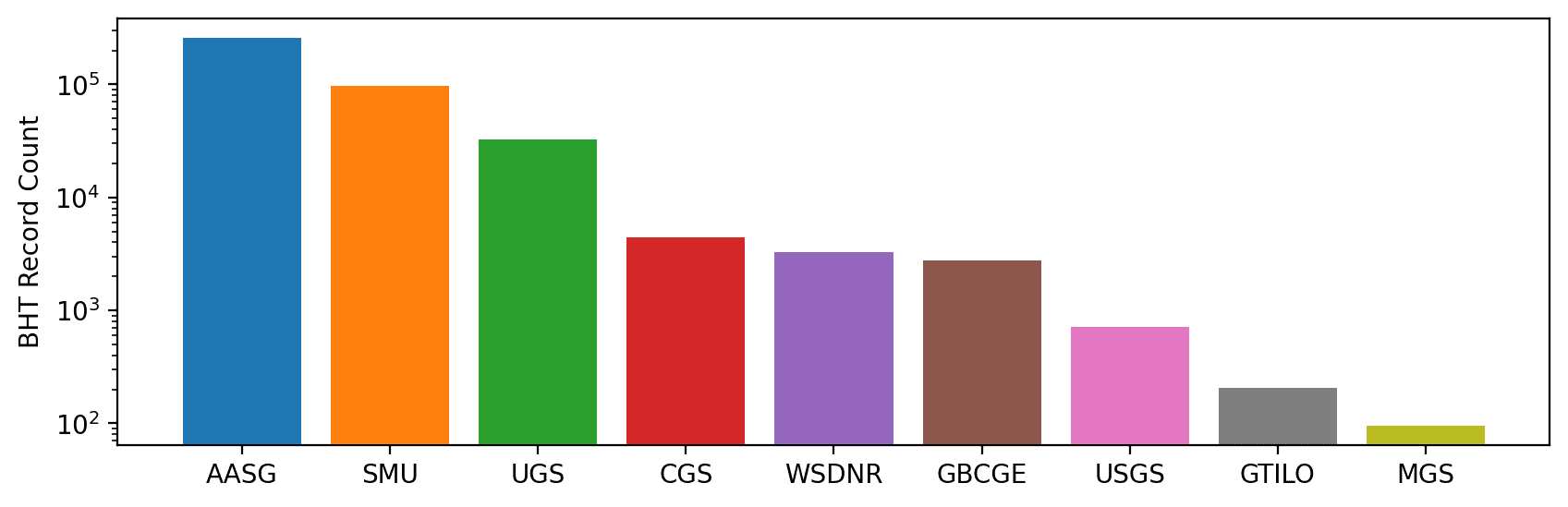}
\caption{Log-scale plot of BHT record count per data source.}\label{fig_bht_hist}
\end{figure}

\begin{figure}[h!]%
\centering
\includegraphics[width=1.0\textwidth]{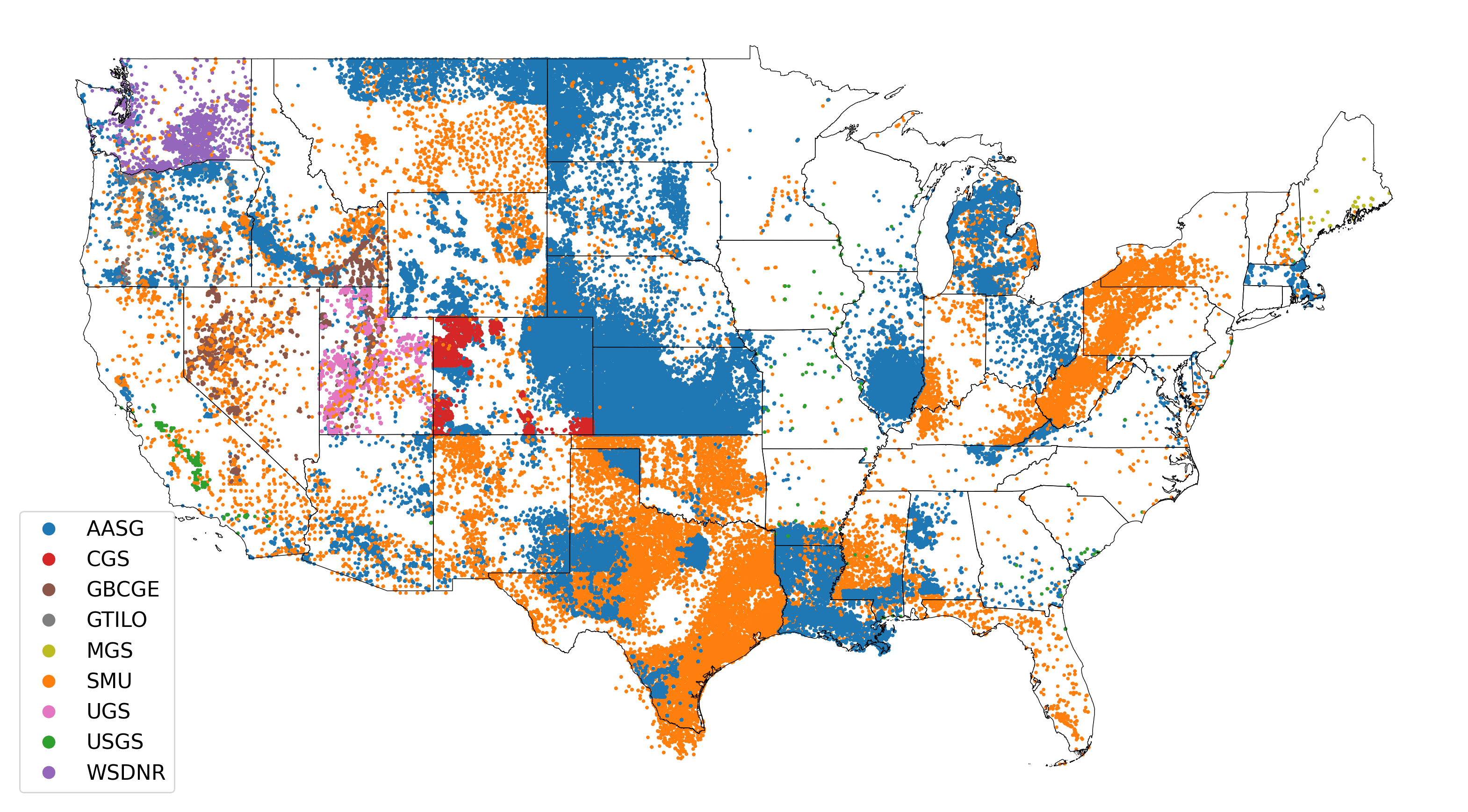}
\caption{BHT records projected on the conterminous US, colored based on data source.}\label{fig_bht_source}
\end{figure}

\begin{figure}[h!]%
\centering
\includegraphics[width=1.0\textwidth]{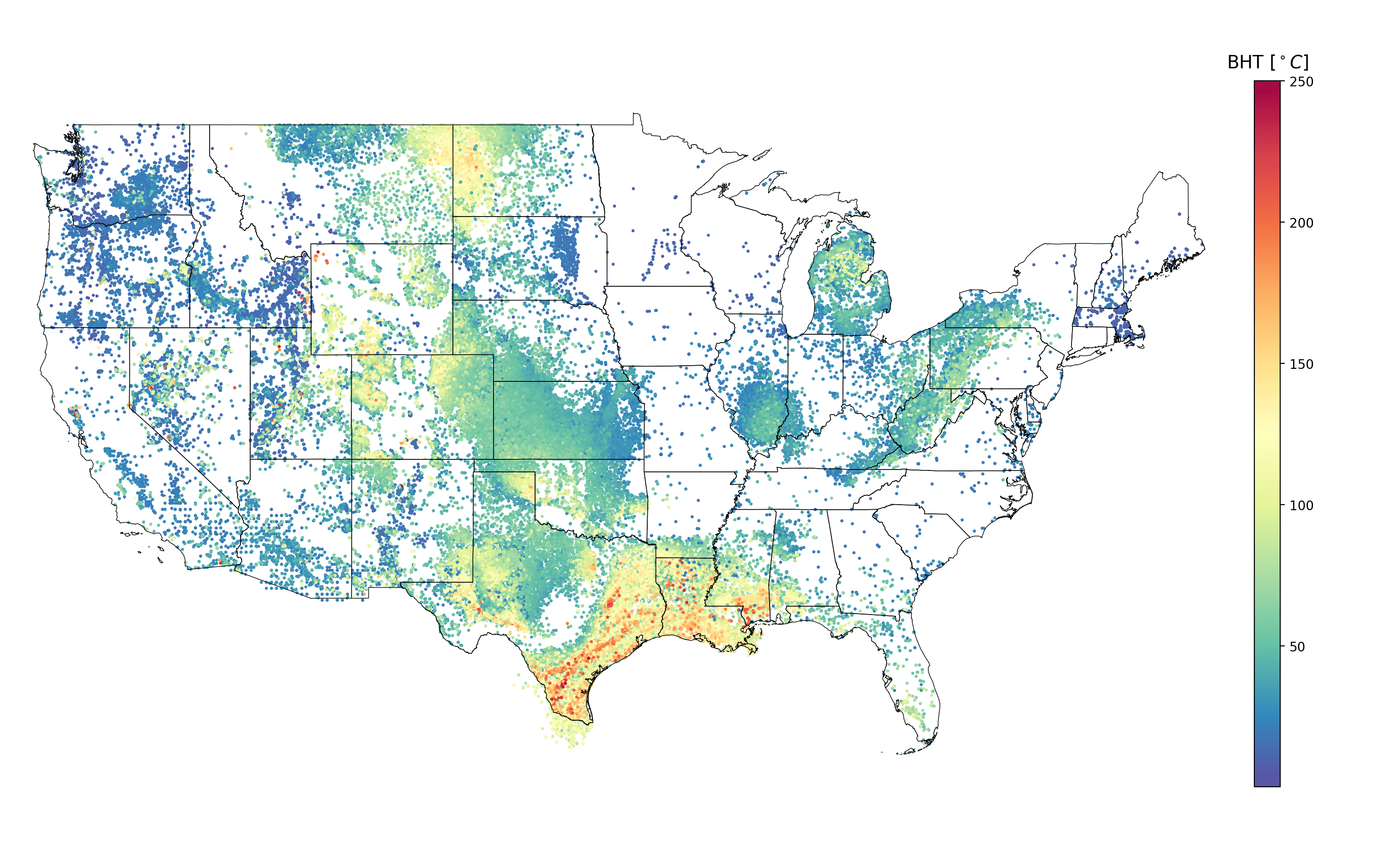}
\caption{BHT records projected on the conterminous US, colored based on magnitude. Note that we set 250$^{\circ} \ C$ as an upper threshold for visual convenience.}\label{fig_bht_magnitude}
\end{figure}

Due to the temperature differences between the rock and the wellbore fluid, those BHT measurements have to be corrected \cite{jessop1990developments}. Due to mud circulation and borehole temperature disequilibrium, this thermal disturbance is not equally distributed along depth. Consequently, compared to the adjacent rock, deeper and shallower sections indicate lower and higher temperatures, respectively \cite{schumacher2020new}. To correct BHT measurements, we used the Harrison correction which is based on a second-order polynomial equation developed by the Oklahoma Geological Survey \cite{harrison1982geothermal}. We found this method to be more suitable to this study as it can be used with only temperature and depth measurements available. Meanwhile, duplicated BHT measurements across the different data sources were detected and filtered out. Note that records are considered to be duplicates if they have the same depth, Easting, and Northing when rounded to the nearest 0.1 m, 0.1 km, and 0.1 km, respectively. As shown in Fig \ref{fig_bht_hist}, our database included a total of 400,134 unique BHT measurements with maximum and minimum corrected BHT measurements of $0.16$ and $533.71^{\circ} \ C$, respectively (going forward, we will drop the term ``corrected'' for brevity). Figs \ref{fig_bht_source} and \ref{fig_bht_magnitude} show all data points projected on the conterminous US and colored based on source and BHT, respectively.

\begin{figure}[h!]%
\centering
\includegraphics[width=1.0\textwidth]{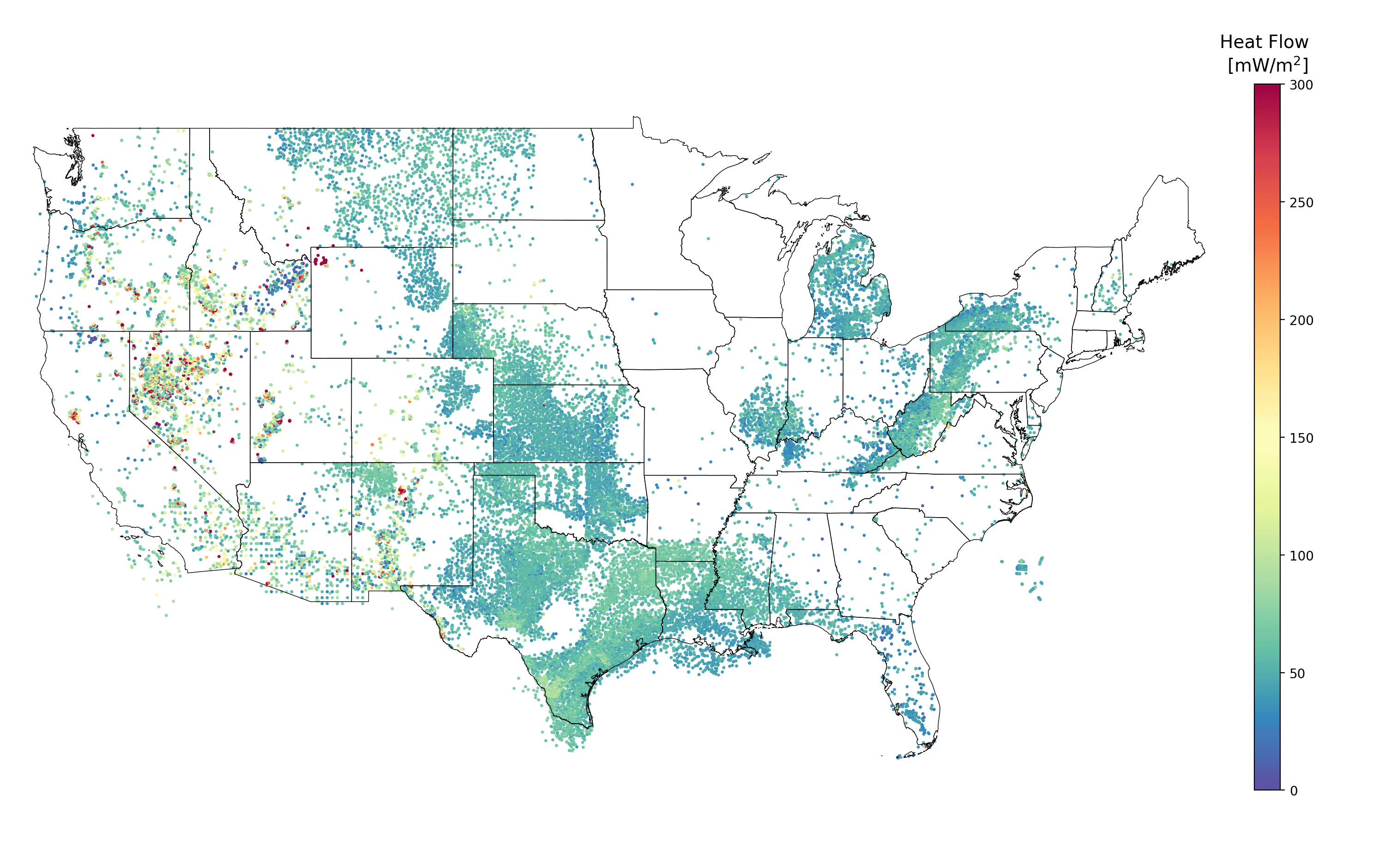}
\caption{Heat flow records for the conterminous US, colored based on magnitude. Note that we set $300 \ mW/m^2$ as an upper threshold for visual convenience.}\label{fig_heat_flow}
\end{figure}

\begin{figure}[h!]%
\centering
\includegraphics[width=1.0\textwidth]{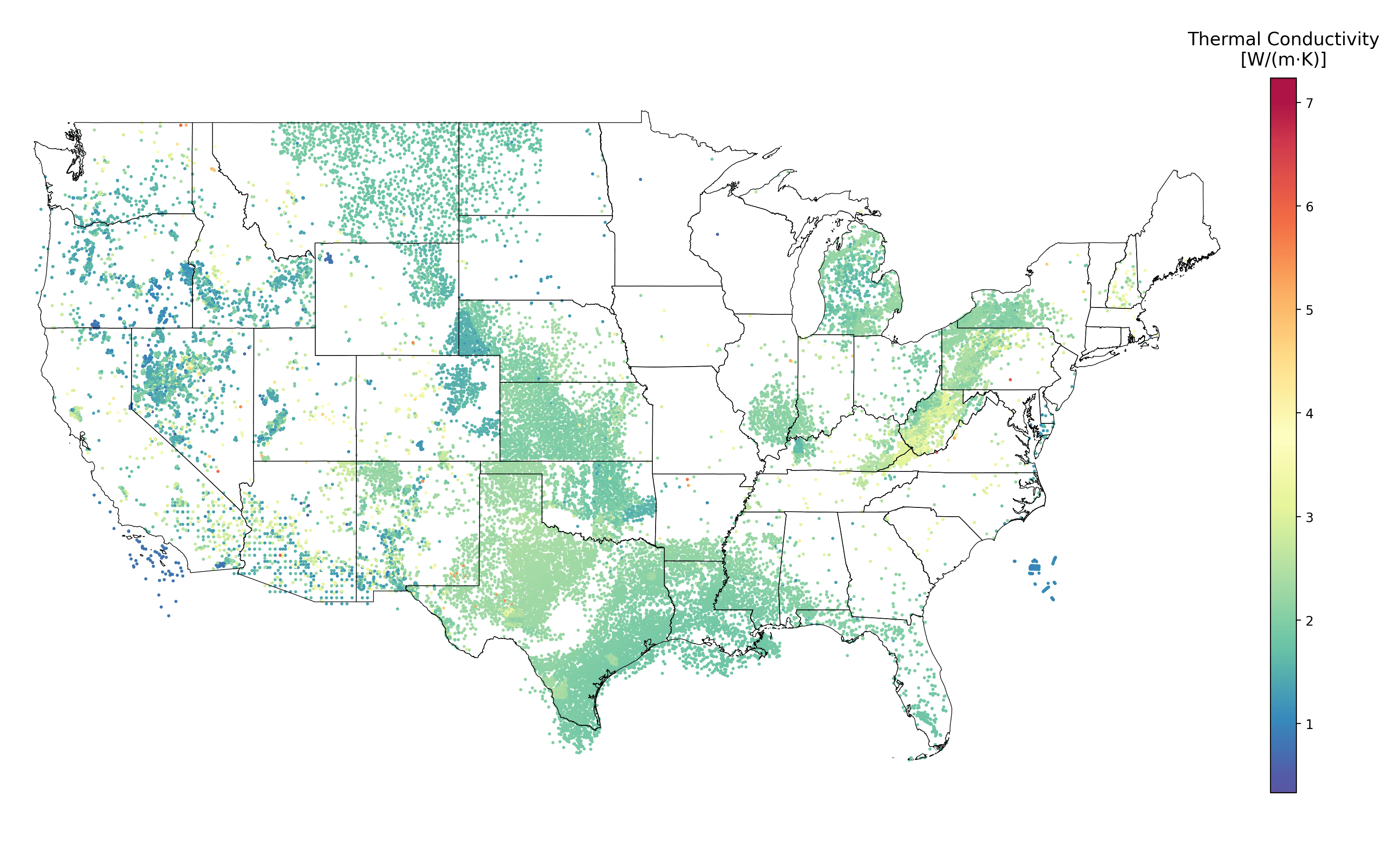}
\caption{Thermal conductivity records for the conterminous US, colored based on magnitude.}\label{fig_thermal_cond}
\end{figure}

The SMU dataset also aggregated data for heat flow and rock thermal conductivity \cite{blackwell2011temperature}. As seen in Fig \ref{fig_heat_flow}, this dataset indicates heat flow estimates as low as nearly zero and as high as $23,012 \ mW/m^2$ at Yellowstone, with median of $59 \ mW/m^2$. Generally, we observed several anomalous heat flow estimates around calderas, such as Yellowstone, Newberry, and Valles. Meanwhile, records indicate rock thermal conductivity values ranging between $0.33$ and $7.24 \ W/m-K$, with median of $2.13 \ W/m-K$, as seen in Fig \ref{fig_thermal_cond}.

\begin{figure}[h!]%
\centering
\includegraphics[width=1.0\textwidth]{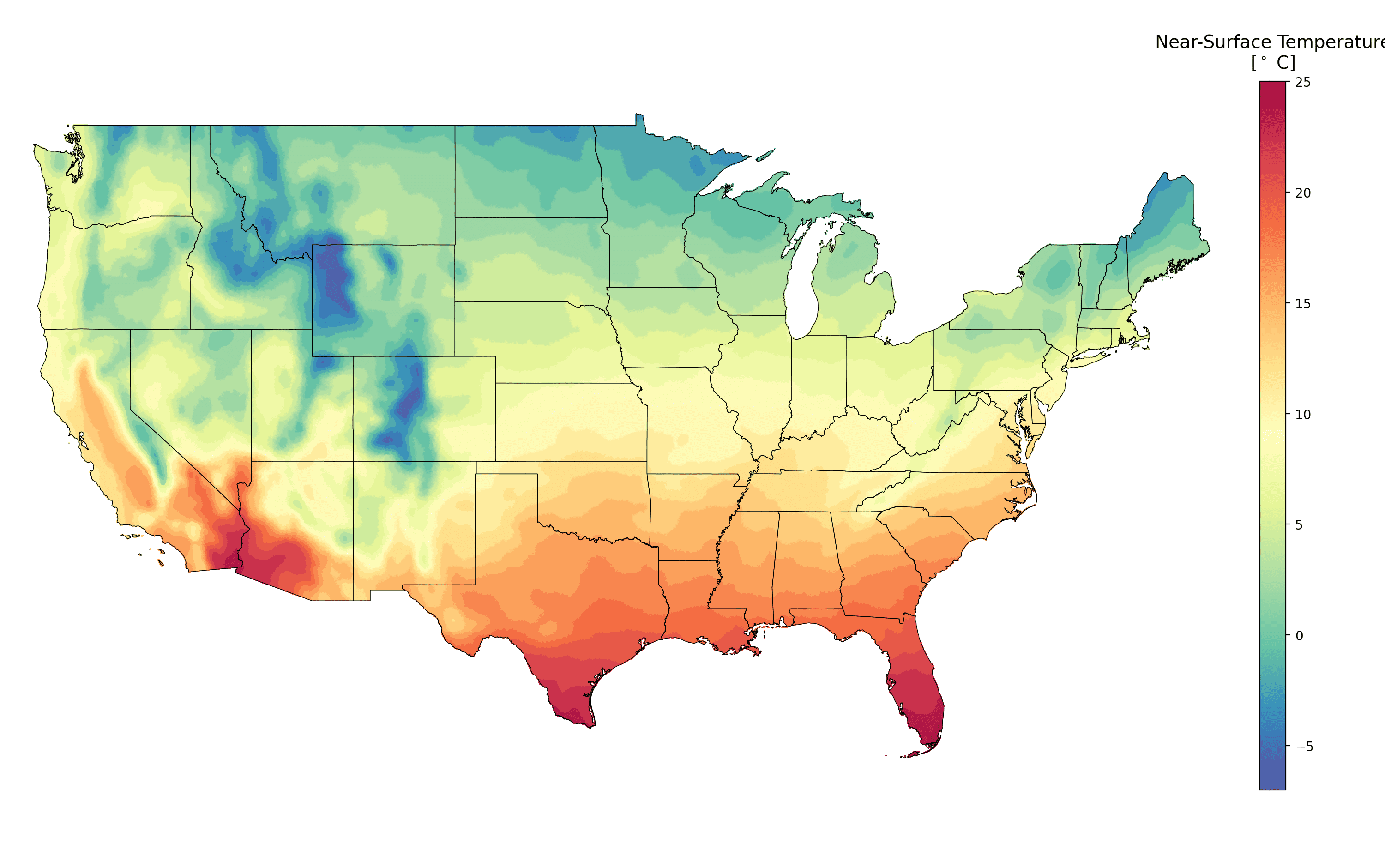}
\caption{Surface temperature records for the conterminous US, colored based on magnitude.}\label{fig_surface_temp}
\end{figure}

Near-surface temperature data is based on average 40-inch soil temperatures reported by the National Weather Service. As seen in Fig \ref{fig_surface_temp}, the average surface temperature ranges between $-7$ and $25^{\circ} \ C$, with mean and standard deviation of $6.5\pm7.6^{\circ} \ C$.

\FloatBarrier
\subsection{Elevation and Sediment Thickness}\label{sub2.2}
\begin{figure}[h!]%
\centering
\includegraphics[width=1.0\textwidth]{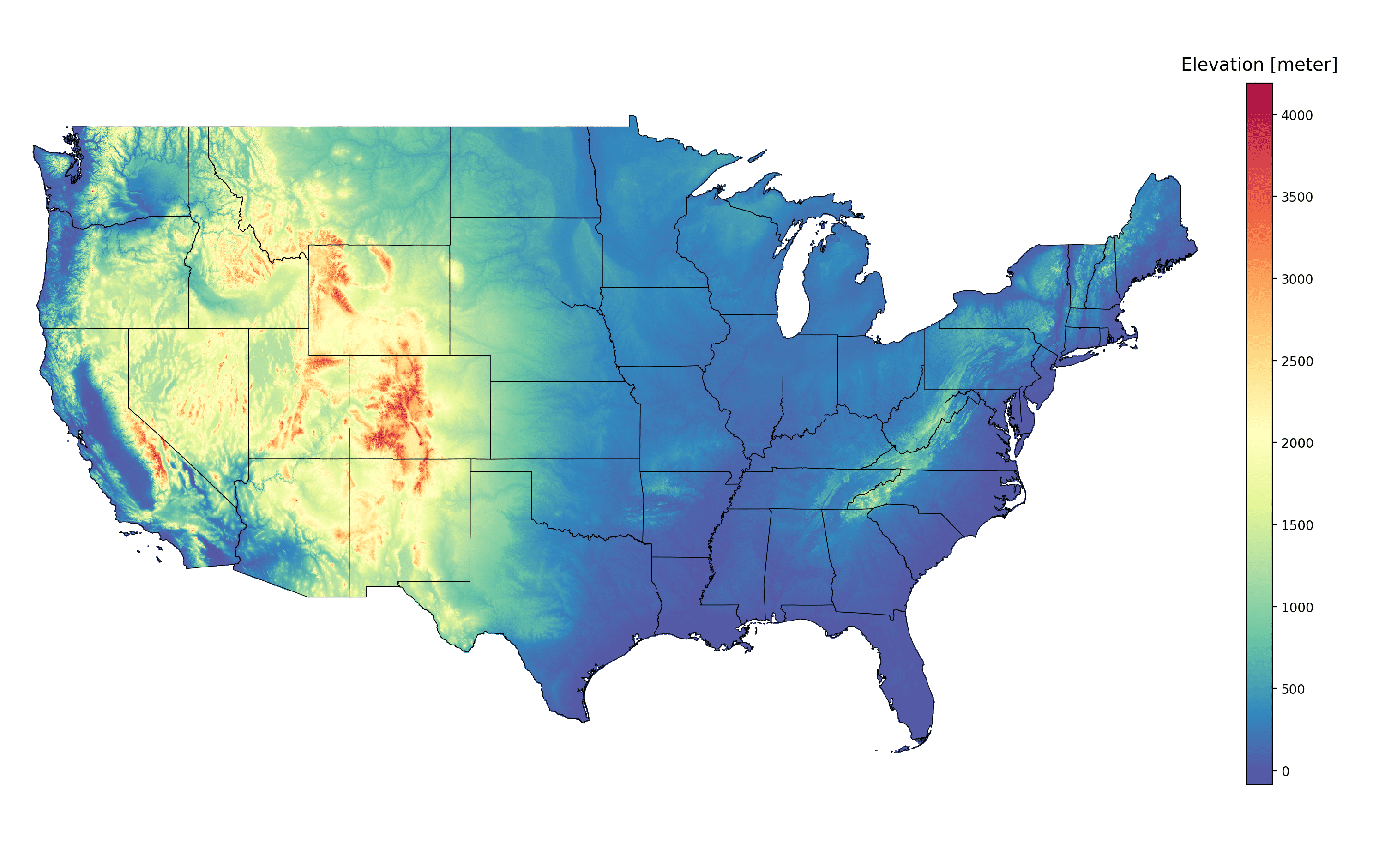}
\caption{Elevation records for the conterminous US, colored based on magnitude.}\label{fig_elevation}
\end{figure}

\begin{figure}[h!]%
\centering
\includegraphics[width=1.0\textwidth]{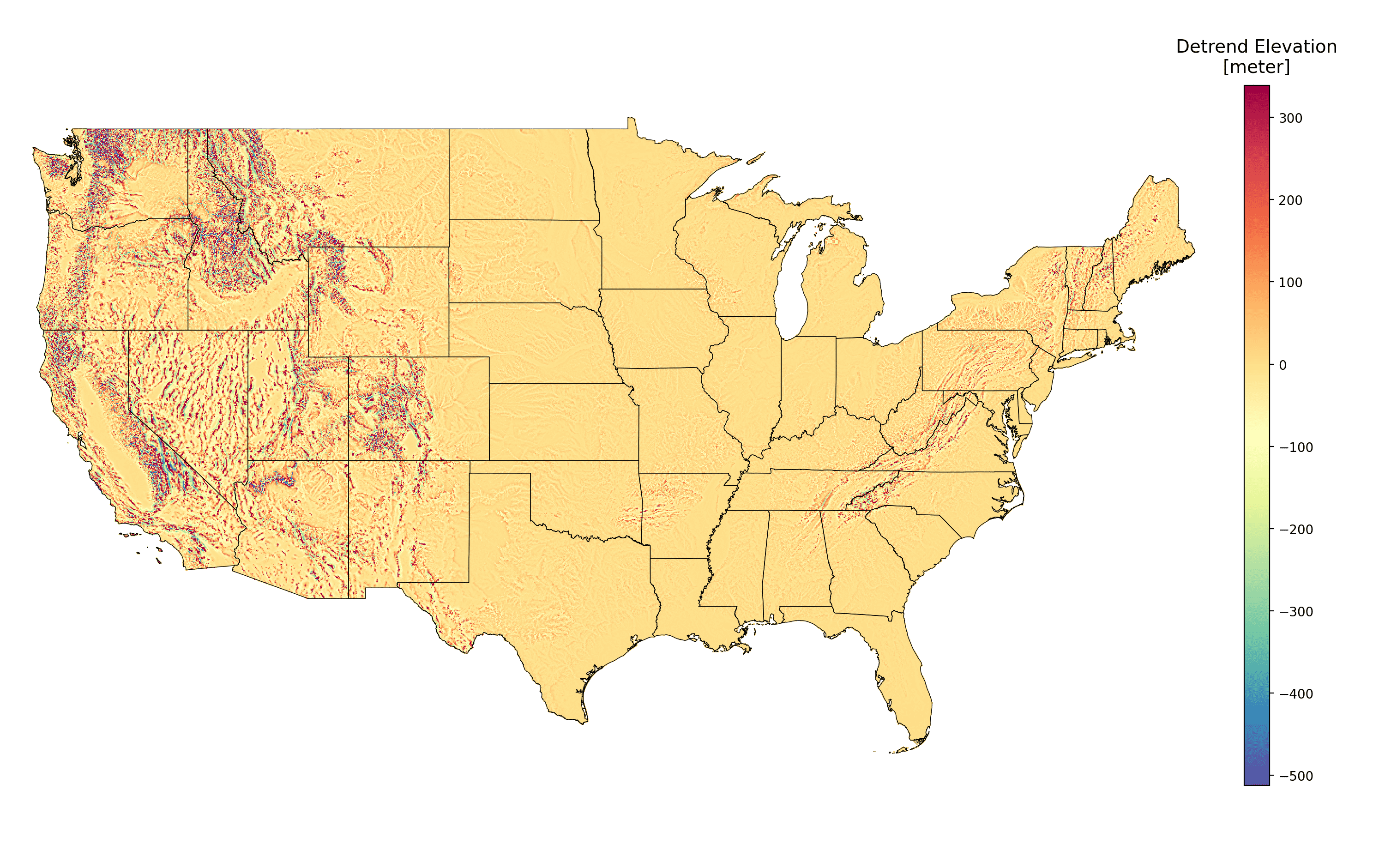}
\caption{Detrended elevation map for the conterminous US, colored based on magnitude.}\label{fig_elevation_detrend}
\end{figure}

\begin{figure}[h!]%
\centering
\includegraphics[width=1.0\textwidth]{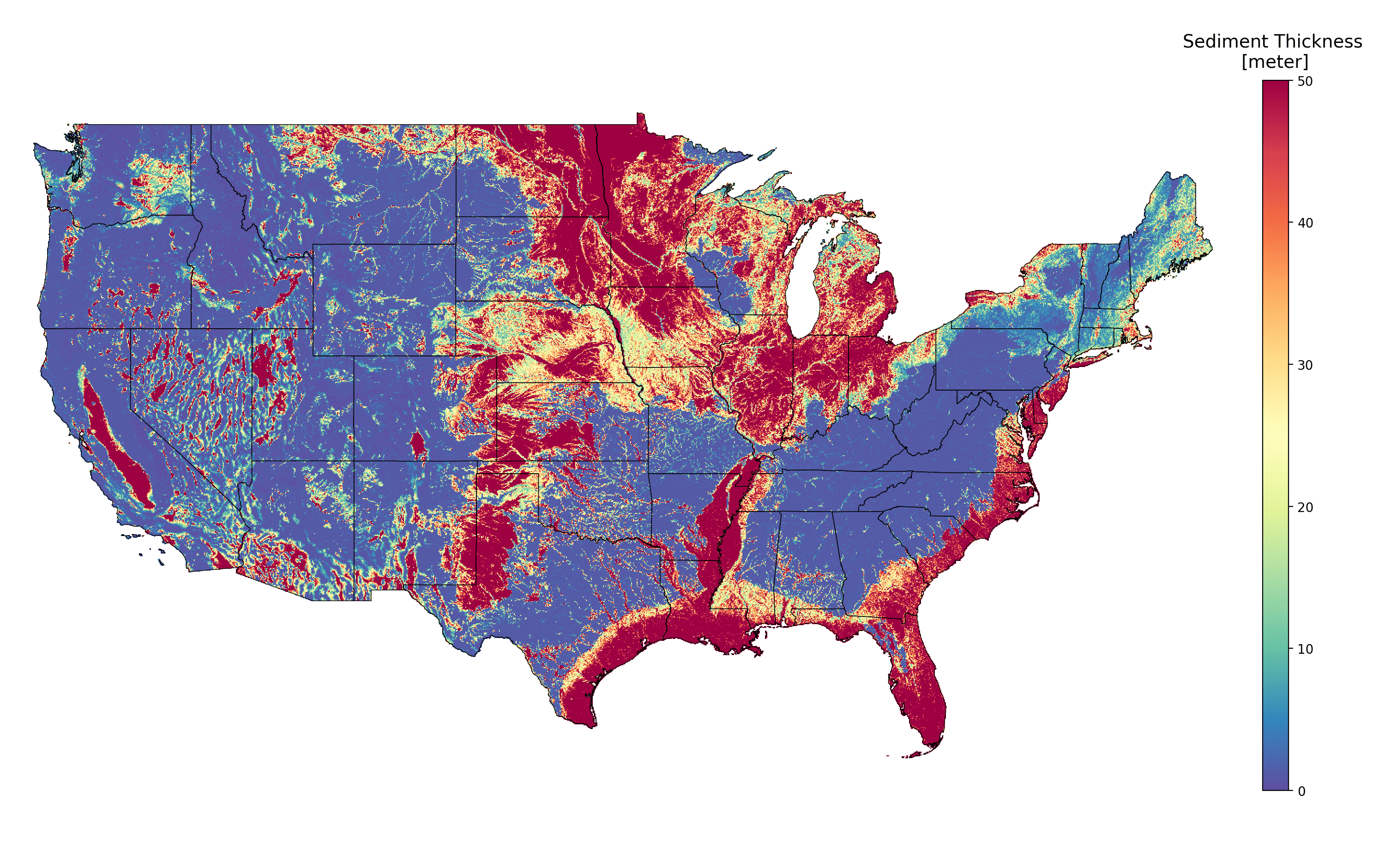}
\caption{Sediment thickness records for the conterminous US, colored based on magnitude. Values of $50 m$ indicate sediment thickness of $\geq 50 m$ while lower values are indicative of exact value estimated by the data source}\label{fig_sed_thick}
\end{figure}

\begin{figure}[h!]%
\centering
\includegraphics[width=1.0\textwidth]{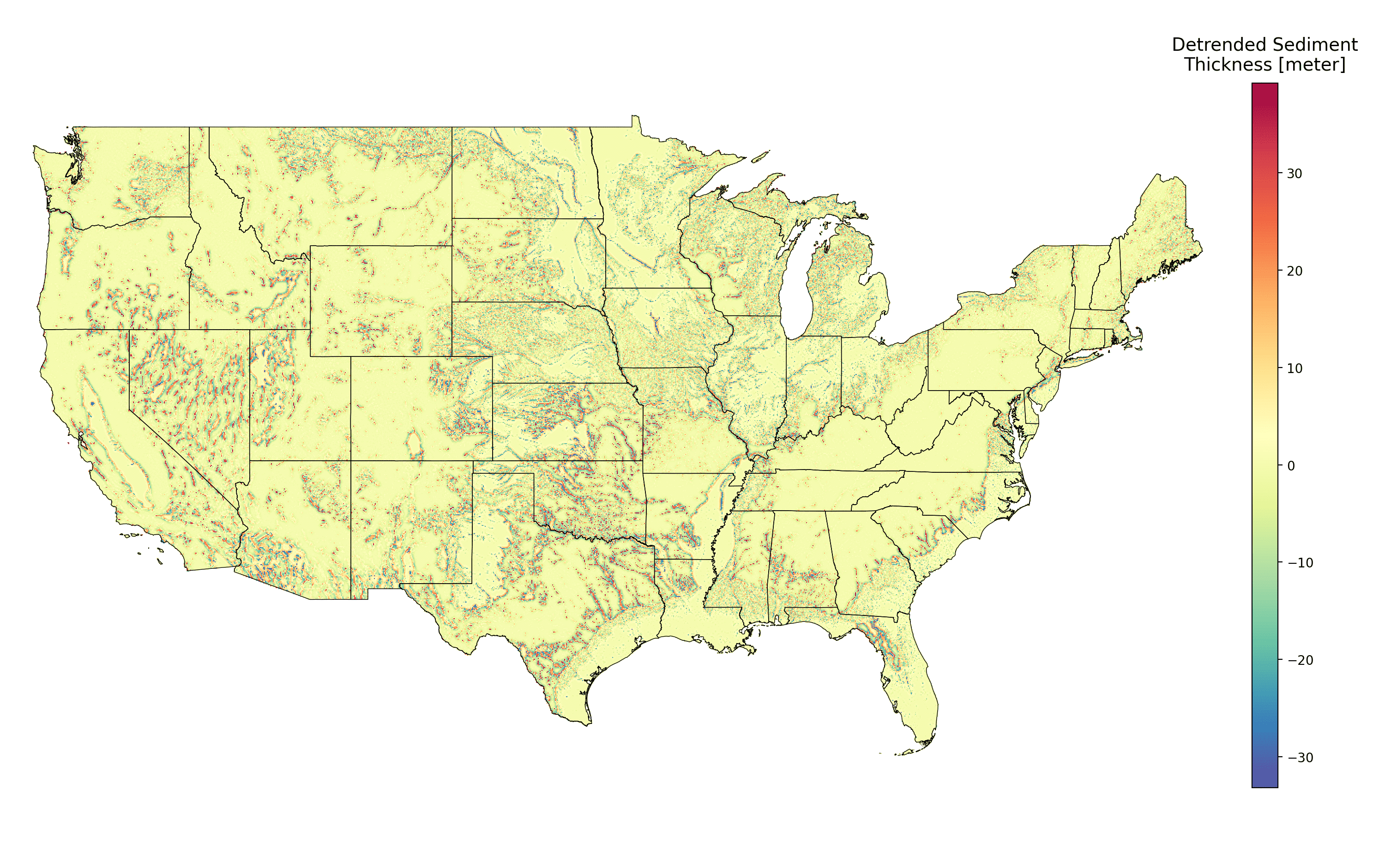}
\caption{Detrended sediment thickness map for the conterminous US, colored based on magnitude.}\label{fig_sediment_thickness_detrend}
\end{figure}

Additionally, we gathered high-resolution maps depicting elevation and sediment thickness across the conterminous US. Elevation data is queried from the National Map's Elevation Point Query Service offered by the USGS, which has an overall resolution of 0.53 meters \cite{arundel2018assimilation}. As seen in Fig \ref{fig_elevation}, elevation ranges between $-84$ and $4,192 \ m$, with mean and standard deviation of $791\pm726 \ m$. Sediment thickness data were acquired through the Oak Ridge National Laboratory Distributed Active Archive Center. This dataset provides estimates of the thickness of the permeable layers above bedrock with spatial resolution of $1 \ km^2$ based on input data for topography, climate, and geology. These data are modeled to represent estimated thicknesses by landform type for the geological present \cite{pelletier2016global}. Seen in Fig \ref{fig_sed_thick}, this dataset has resolution up to $50 \ m$, where values of $50 \ m$ indicate sediment thickness of $\geq 50 \ m$ while lower values are indicative of exact value estimated by the data source. To emphasize the basin-and-ridge scale structural features, we created detrended maps for both elevation and sediment thickness, seen in Figs \ref{fig_elevation_detrend} and \ref{fig_sediment_thickness_detrend}. We followed a simple procedure using a $30 \ km \ \times \ 30 \ km$ moving window to compute a trend map for each quantity which is then subtracted from the corresponding raw map to finally arrive at the detrended map.

\FloatBarrier
\subsection{Magnetic and Gravity Anomalies}\label{sub2.3}
\begin{figure}[h!]%
\centering
\includegraphics[width=1.0\textwidth]{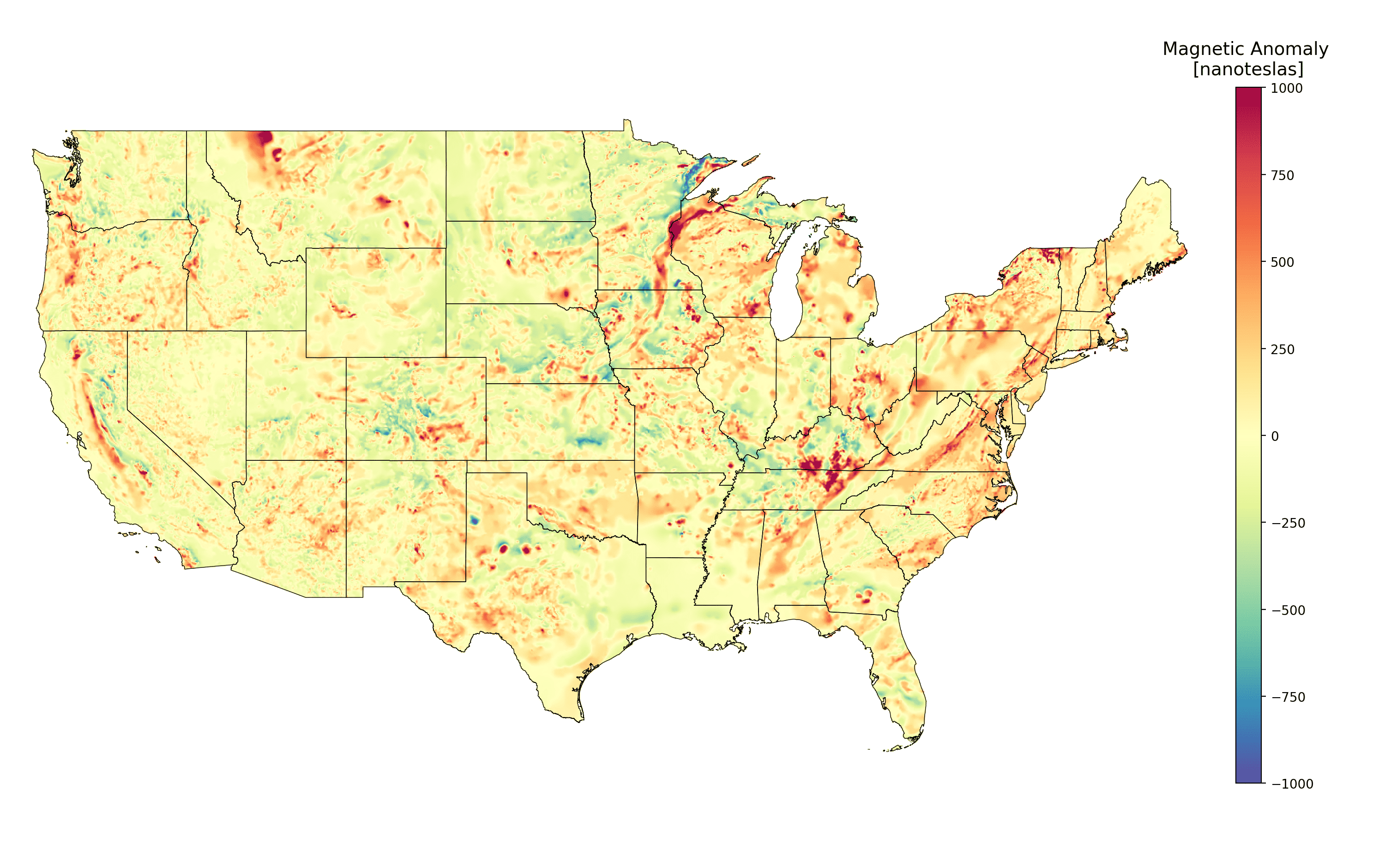}
\caption{Magnetic anomaly for the conterminous US, colored based on magnitude. Note that we set $-1000$ and $1000 \ nanoteslas$ as an lower and upper thresholds, respectively, for visual convenience.}\label{fig_magnetic}
\end{figure}

\begin{figure}[h!]%
\centering
\includegraphics[width=1.0\textwidth]{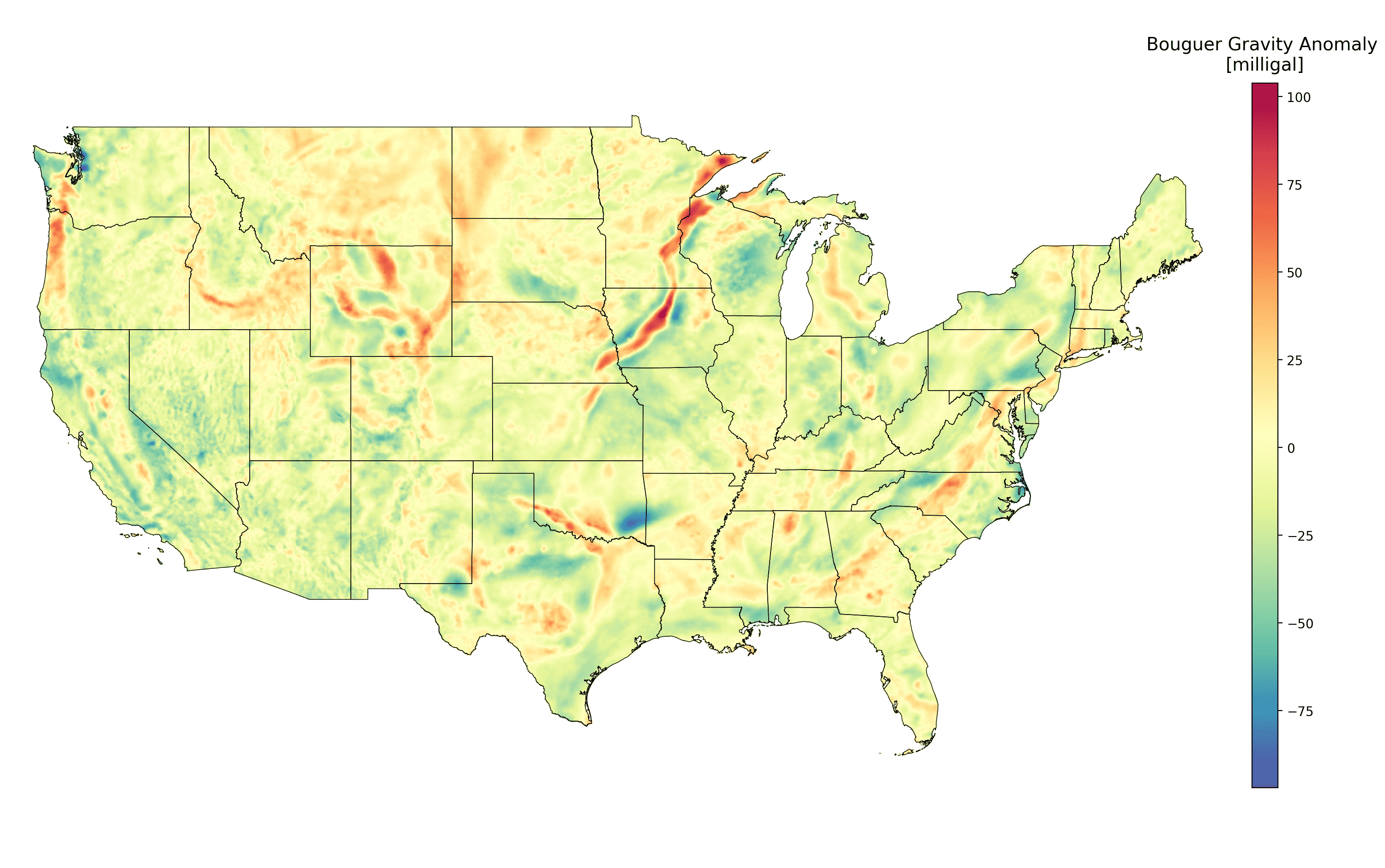}
\caption{Bouguer gravity anomaly for the conterminous US, colored based on magnitude.}\label{fig_gravity}
\end{figure}

Magnetic and gravity anomalies were acquired through the USGS Mineral Resources Online Spatial Data. This data is a result of a joint effort between the USGS, the Geological Survey of Canada, and Consejo de Recursos Minerales of Mexico \cite{bankey2002digital}. The airborne measurement of the earth's magnetic field over North America describes the magnetic anomaly caused by variations in earth materials and structure. We accessed the corrected aeromagnetic grid data with a resolution of $1 \ km^2$ per grid cell. Seen in Fig \ref{fig_magnetic}, magnetic anomalies for the conterminous US range between $-1,513$ to $2,614 \ nanoteslas$, with mean and standard deviation of $41\pm206 \ nanoteslas$. Bouguer gravity anomaly data was based on National Information Mapping Agency data files which were gathered and curated by the USGS with a spatial resolution of nearly $10 \ km^2$ per grid cell. Gravity anomalies are produced by density variations within the rocks of the Earth's crust and upper mantle. The free-air correction reduces the measurement to sea level by assuming there is no intervening mass as a uniform slab of constant density, and the Bouguer correction includes the effects of constant density topography within 166.7 km of the measurement location \cite{kucks1999bouguer}. Seen in Fig \ref{fig_gravity}, Bouguer gravity anomaly ranges between $-224$ and $105 \ milligal$, with mean and standard deviation of $-8.9\pm19.8 \ milligal$.

\FloatBarrier
\subsection{Radioactive Elements}\label{sub2.4}
\begin{figure}[h!]%
\centering
\includegraphics[width=1.0\textwidth]{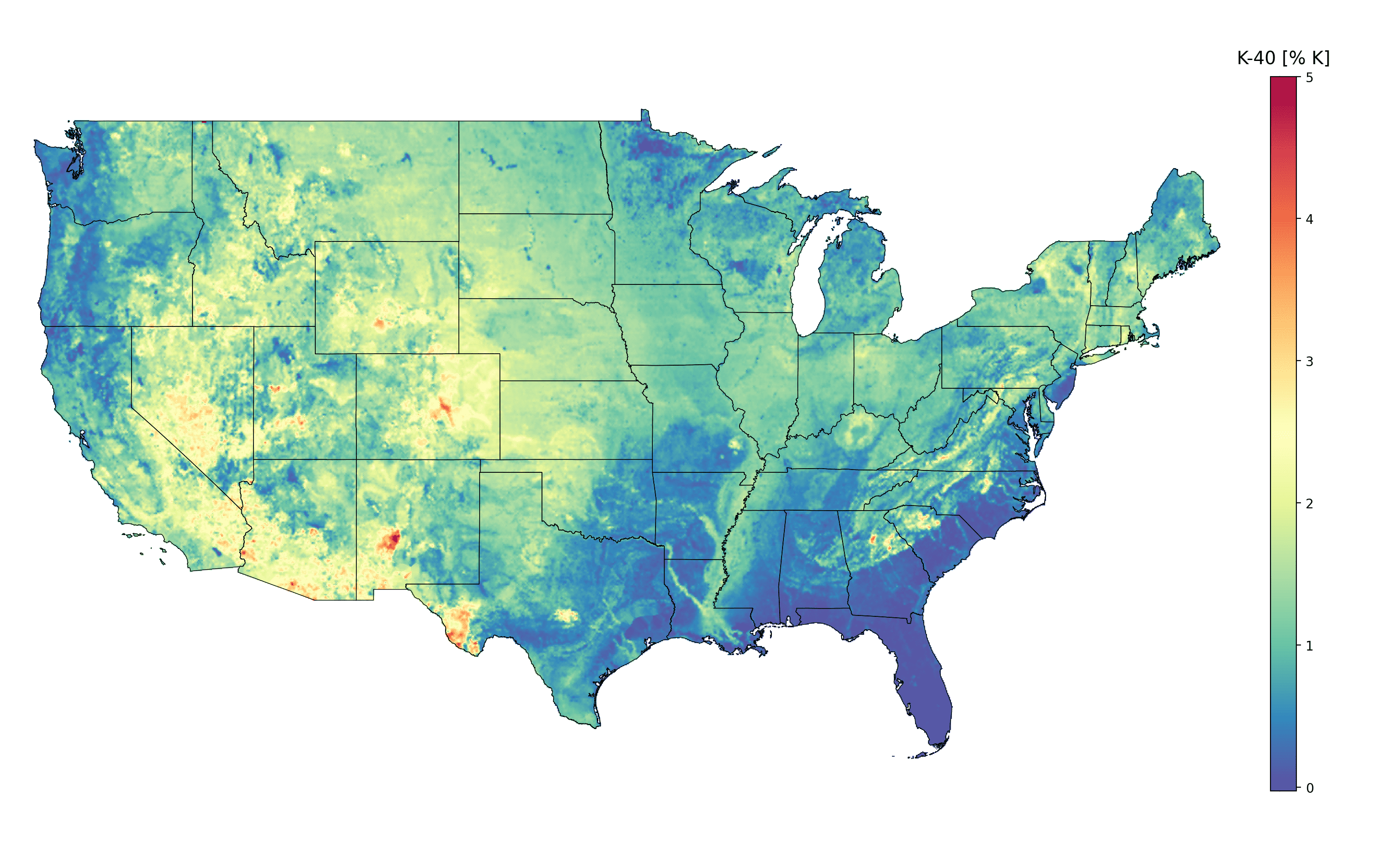}
\caption{K-40 radioactive decay for the conterminous US, colored based on magnitude. Note that we set $5 \ \% K$ as an upper threshold for visual convenience.}\label{fig_k}
\end{figure}

\begin{figure}[h!]%
\centering
\includegraphics[width=1.0\textwidth]{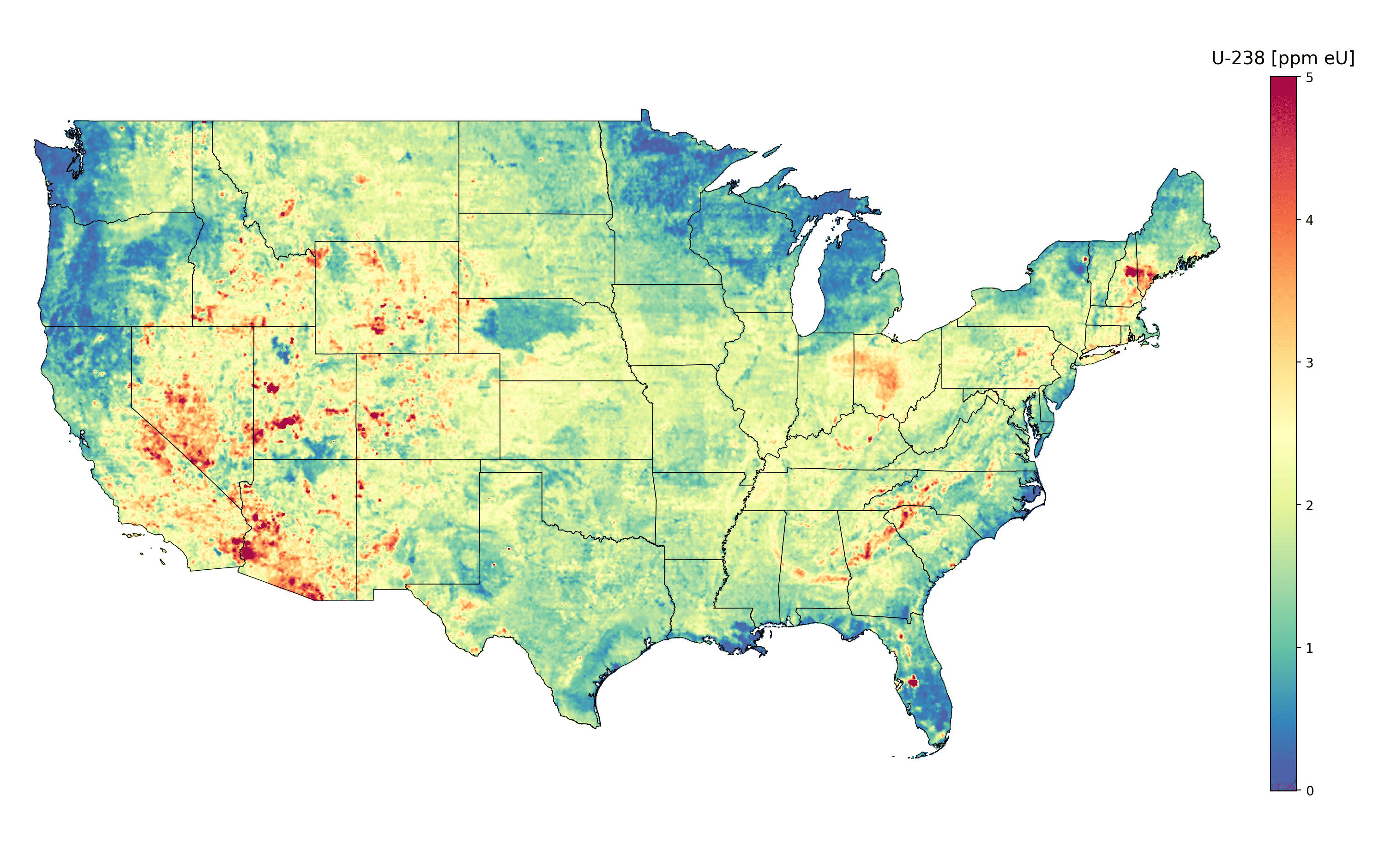}
\caption{U-238 radioactive decay for the conterminous US, colored based on magnitude. Note that we set $5 \ ppm eU$ as an upper threshold for visual convenience.}\label{fig_u}
\end{figure}

\begin{figure}[h!]%
\centering
\includegraphics[width=1.0\textwidth]{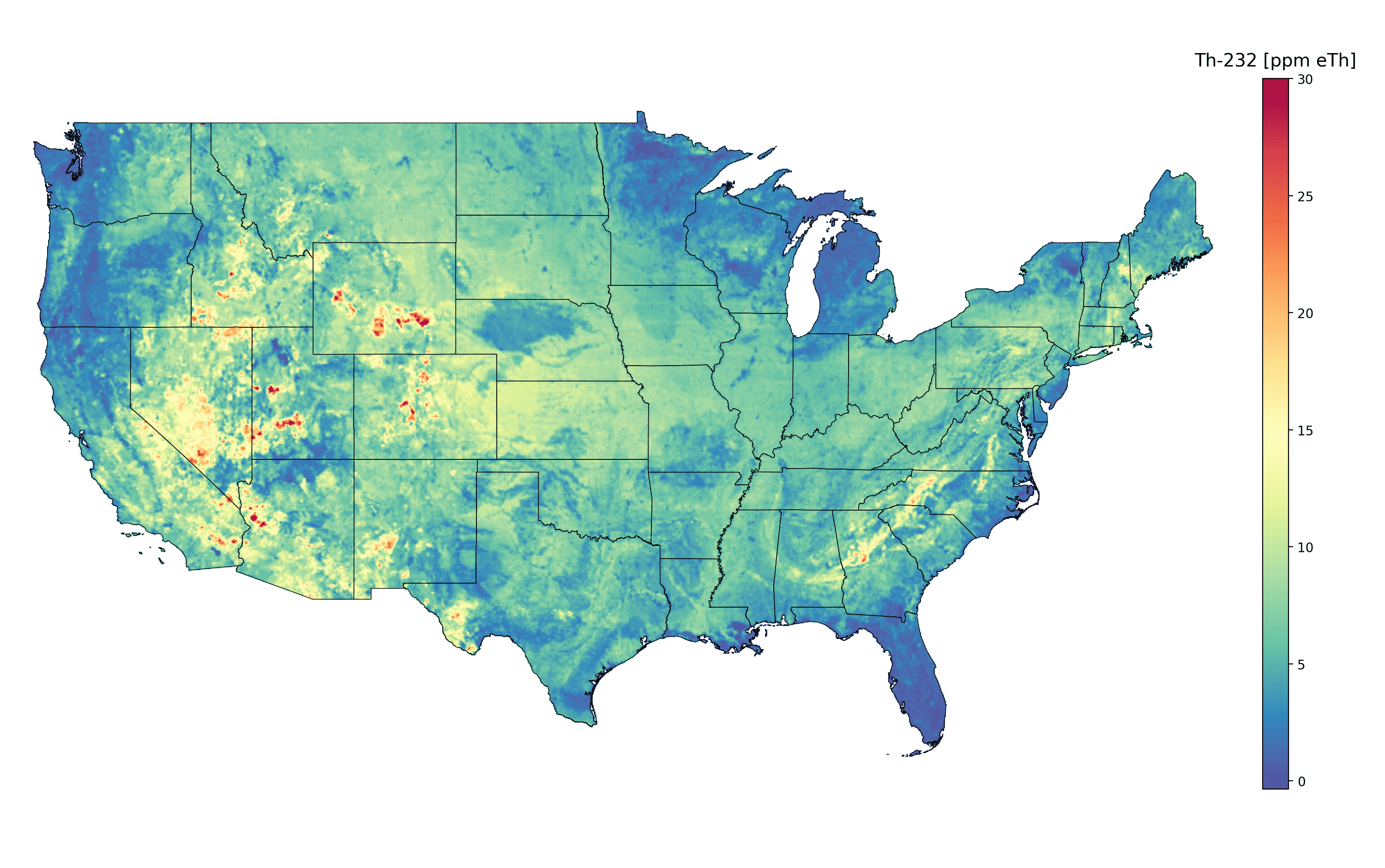}
\caption{Th-232 radioactive decay for the conterminous US, colored based on magnitude. Note that we set $30 \ ppm eTh$ as an upper threshold for visual convenience.}\label{fig_th}
\end{figure}

Aerial gamma-ray surveys measure the gamma-ray flux produced by radioactive decay of the naturally occurring elements K-40, U-238, and Th-232 in the top few centimeters of rock or soil. With a spatial resolution of nearly $2.5 \ km^2$, the USGS aggregated and curated aeroradiometric data for the Conterminous United States which originate to the National Uranium Resource Evaluation Program of the US Department of Energy. \cite{duval2005terrestrial, hill2009aeromagnetic}. Radioactive decay of K-40, U-238, and Th-232 data is reported in the units of percent potassium (\% K), parts per million equivalent thorium (ppm eTh), and parts per million equivalent uranium (ppm eU), respectively. Seen in Figs \ref{fig_k}, \ref{fig_u}, and \ref{fig_th}, radioactive decay of K-40, U-238, and Th-232 shows mean and standard deviation of $1.17\pm1.53 \ \% \ K$, $1.72\pm1.65 \ ppm \ eU$, and $6.19\pm3.64 \ ppm \ eTh$, respectively.

\FloatBarrier
\subsection{Seismicity and Electric Conductivity}\label{sub2.5}
\begin{figure}[h!]%
\centering
\includegraphics[width=1.0\textwidth]{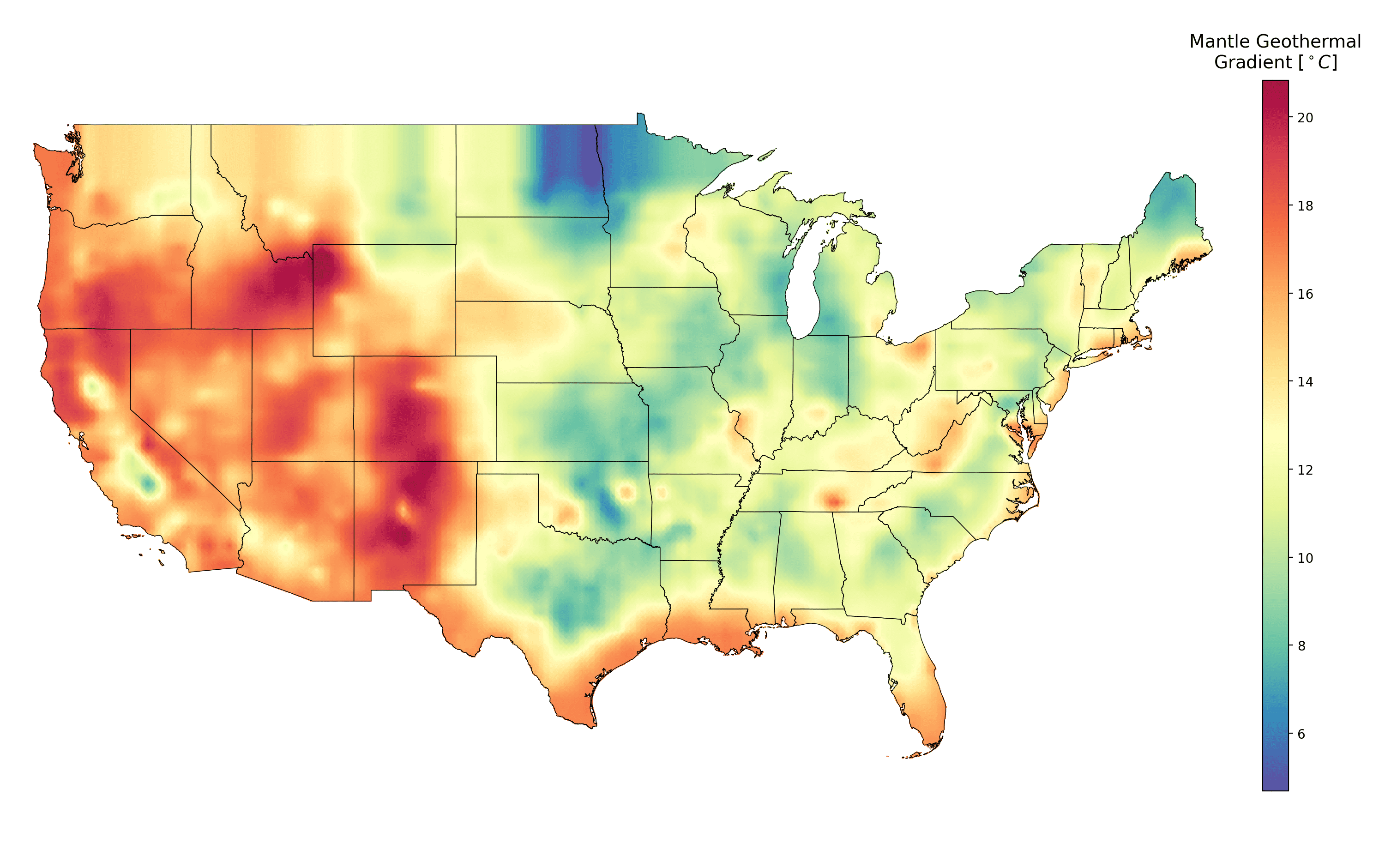}
\caption{Average geothermal gradient at a mantle depth of $60 \ km$ across the continental United States as estimated using seismic measurements.}\label{fig_gg_mantle_60km}
\end{figure}

\begin{figure}[h!]%
\centering
\includegraphics[width=1.0\textwidth]{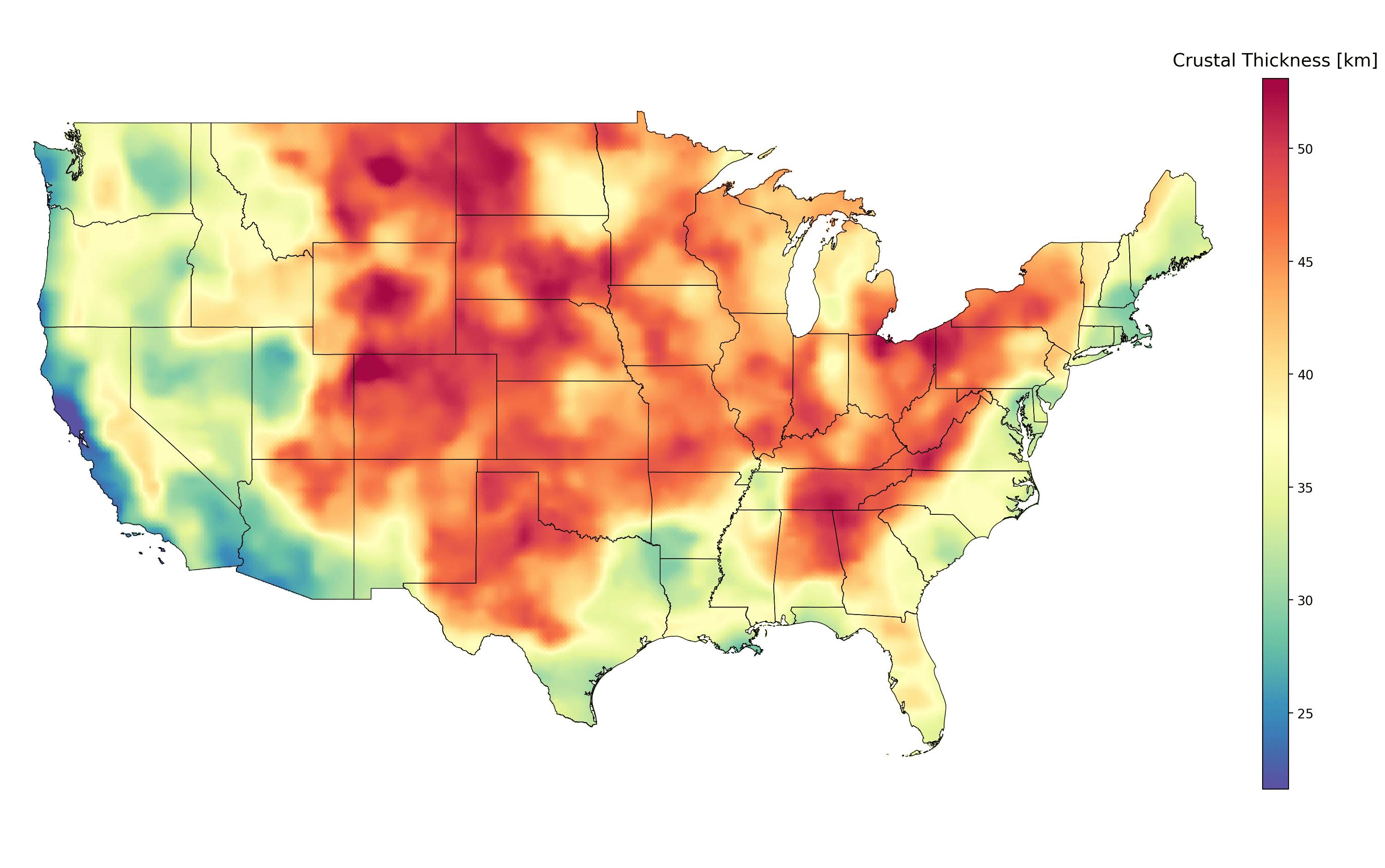}
\caption{Crustal thickness at a depth of $60 \ km$ across the continental United States as estimated using seismic measurements.}\label{fig_crustal_thickness}
\end{figure}

\begin{figure}[h!]%
\centering
\includegraphics[width=1.0\textwidth]{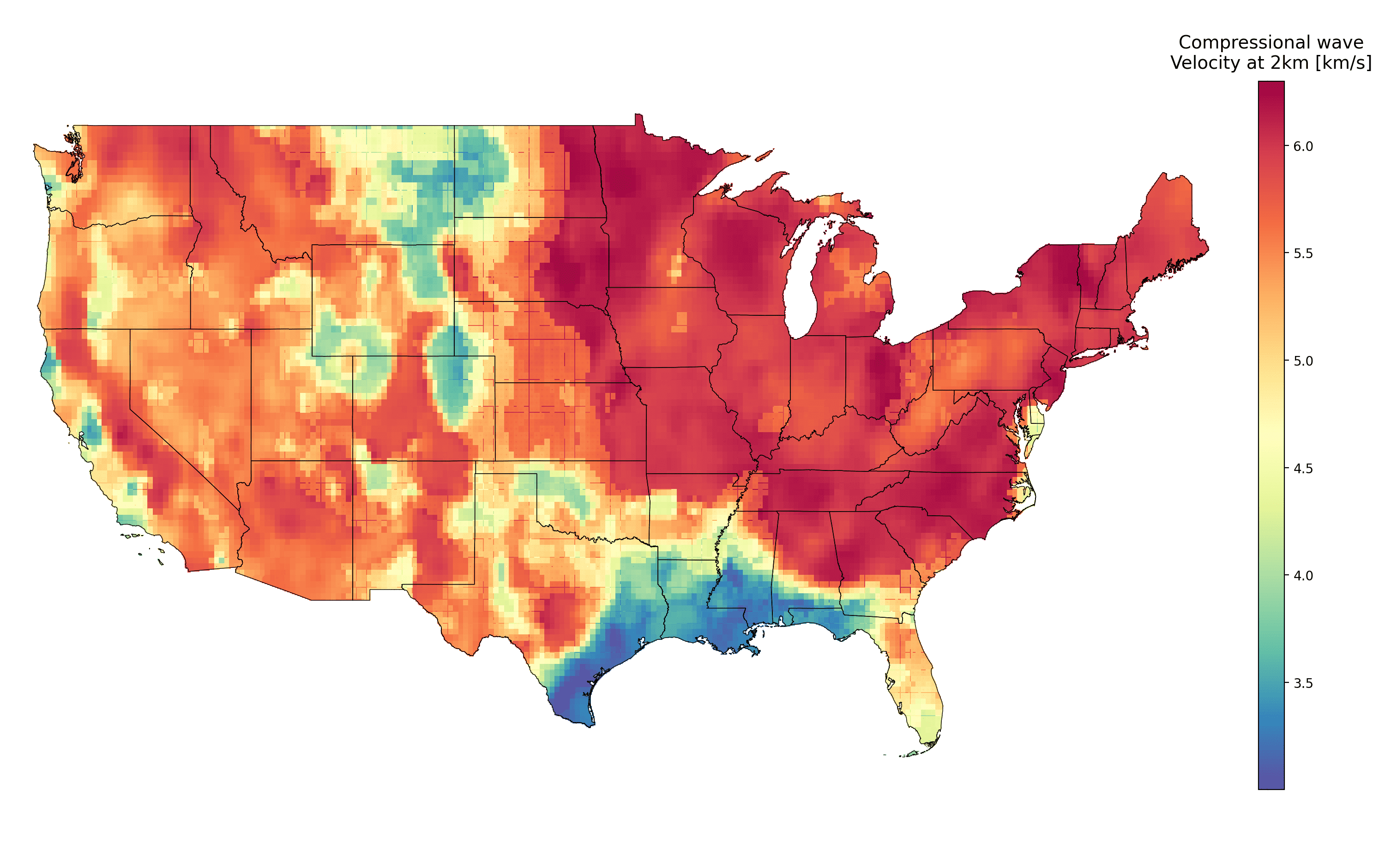}
\caption{Compressional wave velocity at a depth of $2 \ km$ across the continental United States as estimated using seismic measurements.}\label{fig_Vp_2km}
\end{figure}

Seismic wave velocity and electrical conductivity are highly affected by the subsurface rock and fluid properties, and in-situ conditions (e.g., porosity, pore pressure, temperature, elastic moduli). Generally, with increasing in-situ temperatures, seismic wave velocity and thermal conductivity decrease \cite{poletto2018sensitivity, yoshino2011electrical}. We acquired seismic and electric conductivity measurements nation-wide using the \href{http://www.usarray.org/}{USArray project}, which deployed transportable seismic and magnetic stations across the United States over years to passively capture the Earth's seismic activities and magnetic field. We particularly used inverted and processed maps generated by various studies based on data captured by this project, including the average geothermal gradient at the mantle \cite{shinevar2023mantle}, crustal thickness \cite{schmandt2015distinct}, compressional-shear wave velocity ratio \cite{shen2016crustal}, velocity perturbation \cite{golos2020variations}, and electric conductivity \cite{murphy2023geoelectric}. Figs \ref{fig_gg_mantle_60km}, \ref{fig_crustal_thickness}, and \ref{fig_Vp_2km} demonstrate some of these properties after we upsampled them to our target spatial resolution of $18 \ km^2$. Where properties vary with depth (e.g., seismic velocities, velocity ratio, electric conductivity), we additionally normalized each quantity along depth to eliminate their naturally increasing or decreasing trends with depth.

\FloatBarrier
\section{Methods}\label{sec3}
Three-dimensional point clouds are typically unordered, sparse, and unevenly distributed over an irregular grid. They are often associated with point-level features but with various spatial resolutions (e.g., sensor data where measurements are taken at different three-dimensional locations). We are interested in interpolation tasks where a quantity of interest is to be predicted for some desired irregular three-dimensional grid. For instance, our study was interested in regressing temperature-at-depth with accuracy along depth intervals and across the conterminous US.

Popular three-dimensional spatial interpolation techniques include nearest neighbours \cite{rukundo2012nearest}, inverse distance weighing \cite{lu2008adaptive}, kriging \cite{ryu2002kriging}, radial basis function kernels \cite{bhatia2016radial, mo2022implicit}, and spline methods \cite{hutchinson1995interpolating, wahba1990spline}. Each of these methods suffer one or more caveats, e.g., discrete and abrupt surfaces in the presence of dense spatial measurements, smoothing, dismissal of spatial structure information and correlation, linearity, lack of sufficient non-linearity, use of feature engineering and heuristics, and local trends \cite{qiao2019comparison, li2014spatial}. Multiple deep learning architectures were proposed to treat point cloud inputs on irregular grids for various tasks, including classification, regression, segmentation, amongst others. They can be lumped into three categories: (1) point-based methods (e.g., PointNet, PointNet++, PointWeb) \cite{qi2017pointnet, qi2017pointnet++, zhao2019pointweb}, (2) convolution-based methods (e.g., Relation-shape CNN, DensePoint, Pointwise-CNN) \cite{liu2019relation, liu2019densepoint, hua2018pointwise}, (3) registration methods (e.g., PPFNet33, FoldNet, PointNetLK) \cite{besl1992method, deng2018ppfnet, deng2018ppf, gojcic2019perfect, choy2019fully, aoki2019pointnetlk}, and (4) graph-based methods (e.g., Edge-Conditioned Convolution, EdgeConv, 3DGCN) \cite{simonovsky2017dynamic, wang2019dynamic, lin2020convolution}. Whereas these architectures can input irregular three-dimensional point clouds, they are not particularly designed for interpolation where the target quantity can explicitly inform predictions at neighbouring grid cells. Additionally, the presence of extensive sparsity in point cloud data makes it challenging to generalize spatially. We treated these challenges using a novel \textbf{I}nterpolative \textbf{P}hysics-\textbf{I}nformed \textbf{G}raph \textbf{N}eural \textbf{N}etwork (\textbf{InterPIGNN}). The interpolative aspect of this architecture incorporates a graph-based operator which, unlike convolutional neural networks, does not require input data to lie on regular grids. Graph-based architectures are capable of extracting point cloud features without the need for heuristically projecting them on a regular grid. Meanwhile, the physics-informed aspect involves satisfying physical laws related to the problem of interest. In this section, we describe the InterPIGNN module along with the data preprocessing and modeling steps. We conveniently use the three-dimensional temperature-at-depth mapping task as an example interpolation task to further motivate the InterPIGNN module.

\subsection{Interpolative Convolutional Module}\label{sub3.1}
In three-dimensional point cloud interpolation tasks across space of interest, $\Omega$, a target quantity $y$ is measured at a sparsely distributed finite set of points, which we denote as the set of anchor nodes $V_a$ with $|V_a| = n_a$. The objective is to predict or interpolate the target quantity at a finite set of desired points (regular or irregular) sampled from $\Omega$, which we denote as the set of grid nodes $V_g$ with with $|V_g| = n_g$. We denote the union set of the anchor and grid nodes as $V = V_a \bigcup V_g$ with $|V| = n_a + n_g = n$. Each node $v \in V$ is associated with a position vector $\boldsymbol{p_v} \in \Omega$ (in three-dimensional point clouds, $\Omega = \mathbb{R}^3$). We consider two sets of features: (1) high-resolution features $F_{hr}$ which are known at all nodes $v \in \mathcal{V}$, and (2) low-resolution features $F_{lr}$ which are only known at anchor nodes $v \in V_a$. For a node $v \in V$, the high-resolution and low-resolution feature vectors are denoted as $\boldsymbol{x}_{v,hr}$ and $\boldsymbol{x}_{v,lr}$, respectively, where the latter is only defined when $v \in V_a$.

We convert the point cloud into a directed graph representation $G = (V, E)$ by constructing a set of directed edges $E \subseteq V_a \times V$ with $|E| = m$. A directed edge $e_{uv}$ is constructed between nodes $u \in V_a$ and $v \in V$ if $\boldsymbol{p}_u$ belongs to (1) the $k$-nearest neighbours of $\boldsymbol{p}_v$, and/or (2) a sphere centered at $\boldsymbol{p}_v$ with radius $r$, where $k$ and $r$ are hyperparameters. An edge $e_{uv}$ has edge weight $w_{uv}$ defined as the inverse euclidean distance and normalized such that $\sum_{u \in \mathcal{N}(v)} w_{uv} = 1$, where $\mathcal{N}(v)$ is the set of neighbours of node $v$. 

Following the GNN framework of message passing, aggregation, and combination, we define our InterPIGNN convolutional module which can be used flexibly as a nonlinear preprocesser or transformer in any deep learning algorithm inputting point clouds. Consider an GNN architecture based on InterPIGNN with $L$ neural layers. At layer $l=0$, the hidden state of each node $v \in V$ is initialized using the high-resolution feature vector, $\boldsymbol{h}_v^{(0)} = \boldsymbol{x}_{v,hr}$. We also assign each edge $e_{uv}$ an attribute $\boldsymbol{a}_{uv}^{(l)}
= \boldsymbol{h}_u^{(l)} - \boldsymbol{h}_v^{(l)}$ that is updated across convolutional layers. For a target node $v \in V$, the InterPIGNN module, seen in Equation \ref{eq-internet}, consists of four neural networks: (1) root network $f_{\theta_r}$ to convolve root high-resolution features, (2) neighbour network $f_{\theta_n}$ to transform each individual message passed by neighbouring nodes including features, positions, edge attributes, and labels, and (3) global network $f_{\theta_g}$ to transform aggregated messages, and (4) post-processing network $f_{\theta_p}$ to jointly transform the outputs of the root and global networks. Note that $\parallel$ represents the vector concatenation operator. Also, note that permutation-invariant aggregation is achieved in the max pooling operation. The final hidden $\boldsymbol{h}_v^{(L)}$ for each node $v \in V$ can then be used for the desired task.

\begin{equation}
\boldsymbol{h}_v^{(l+1)}
= f_{\theta_p} \Bigg[f_{\theta_r}(\boldsymbol{h}_v^{(l)})
+ f_{\theta_g}\Bigg(\frac{1}{|\mathcal{N}(v)|} \sum_{u \in \mathcal{N}(v)} 
\Big[w_{uv} \cdot f_{\theta_n}(\boldsymbol{x}_u \parallel \boldsymbol{p}_u \parallel \boldsymbol{a}_{uv}^{(l)} \parallel y_u)\Big]\Bigg)\Bigg]
\label{eq-internet}
\end{equation}

\subsection{Three-Dimensional Heat Conduction}\label{sub3.2}
Temperature within the Earth's subsurface is governed by a complex interplay of physical processes, mainly thermal conduction, which drives heat transfer through the subsurface materials. A significant contribution also comes from the radioactive decay of isotopes such as uranium, thorium, and potassium, which generates heat. In addition, convective movements of magma and groundwater can redistribute heat vertically and horizontally through advection, altering temperature profiles. The latent heat associated with phase changes of water and tectonic activities, such as volcanic upwelling, further influence subsurface temperatures \cite{jaupart20077}. Collectively, these factors establish the thermal state of the subsurface, varying significantly with geographic location, depth, and local geophysical and hydrological conditions.

Using the collected data for this study, we could satisfy the steady-state three-dimensional heat conduction in the Earth's subsurface which is represented by a partial differential equation (PDE) as seen in Eqs. \ref{eq-conduction}-\ref{eq-heatflow} \cite{cengel2012ebook, mareschal2013radiogenic}. Subsurface temperature $T$, rock thermal conductivity $K$, crustal heat flow $Q_c$, and local heat generation $\dot{g}$ vary with spatially according to easting $x$, northing $y$ and depth $z$. Meanwhile, surface heat flow $Q$ is only a function of $x$ and $y$.

\begin{equation}
\frac{\partial}{\partial x} \Big(K \frac{\partial T}{\partial x}\Big)
+ \frac{\partial}{\partial y} \Big(K \frac{\partial T}{\partial y}\Big)
+ \frac{\partial}{\partial z} \Big(K \frac{\partial T}{\partial z}\Big)
+ \dot{g}
= 0
\label{eq-conduction}
\end{equation}

\begin{equation}
Q_c(z) = Q - \int_{0}^{z} \dot{g} \ dz'
\label{eq-heatflow}
\end{equation}

\subsection{Physics-Informed Graph Neural Networks}\label{sub3.3}

Considering the unique BHT measurements, seen in Fig \ref{fig_bht_magnitude}, the temperature-at-depth interpolation task is a three-dimensional setting with $n_a = 400,134$ anchor nodes. We aim to map temperature-at-depth from surface and down to $7 \ km$ for the conterminous US with $18 \ km^2$ and $1 \ km$ spatial and depth-wise resolutions, hence we have a total of $n_g = 4,279,536$ grid points. Edges were constructed following the $k$-nearest neighbours strategy with $k=5$. The three-dimensional position vector $\boldsymbol{p}$ was defined using depth, Easting, and Northing. Node labels $y$ were set as the BHT measurement at anchor nodes, but they are unknown at grid nodes. Low-resolution features $F_{lr}$ are BHT, heat flow, and thermal conductivity, depth, northing, easting, sediment thickness, magnetic anomaly, Bouguer gravity anomaly, K-40, U-238, and Th-232 radioactive decay measurements, and seismic quantities. High-resolution features $F_{hr}$ are elevation, sediment thickness, near-surface temperature, magnetic anomaly, Bouguer gravity anomaly, and K-40, U-238, and Th-232 radioactive decay measurements.

Anchor nodes were split into $80\%$, $10\%$, and $10\%$ for training, validation, and testing, respectively. To decorrelate spatial features and ensure proper model evaluation, we split data using three-dimensional cubic blocks, each with length of $10 / km$. We used Feature vectors, both $\boldsymbol{x}_{v,lr}$ and $\boldsymbol{x}_{v,hr}$, were normalized using the training split mean $\boldsymbol{\mu}$ and standard deviation $\boldsymbol{\sigma}$ statistics, seen in Equation \ref{eq-normalize}. Toward satisfying the three-dimensional conductive heat transfer laws, we created three graph neural networks (GNNs) $T_{\theta}$, $Q_{\theta}$, and $K_{\theta}$ to predict subsurface temperature, heat flow, and thermal conductivity simultaneously. Each of these networks utilized the interpolative convolutional module where $f_{\theta_r}$, $f_{\theta_n}$, $f_{\theta_g}$, and $f_{\theta_p}$ were each chosen as two-layer perceptrons with 128 and 64 neurons. To ensure proper first and second gradient flow when training InterPIGNN, we used $Mish$ as an activation function. It has an unbounded positive domain which alleviates issues of gradient saturation, bounded negative domain which enhances generalization, and a continuously differentiable range \cite{misra2019mish}. Moreover, we incorporated two types of dropouts: (1) traditional network neuron dropout of $10\%$ to encourage generalization across input samples \cite{srivastava2014dropout}, and (2) graph edge dropout of $1\%$ to encourage generalization across neighbouring nodes \cite{papp2021dropgnn}.

Whereas $Q_{\theta}$ and $K_{\theta}$ were each designed to output a quantity each, i.e., $\hat{Q}$ and $\hat{K}$, respectively, the $T_{\theta}$ architecture involved a final node-level feedforward neural layer to output two quantities: average surface-to-depth geothermal gradient $\hat{g}$, and near-surface temperature correction $\Delta\hat{T}_0$. Consequently, the predicted subsurface temperature at depth $z$ was computed as $\hat{T}(z) = T_0 + \Delta\hat{T}_0 + \hat{g} \cdot z$, where $T_0$ is the surface temperature seen in Fig. \ref{fig_surface_temp}. We also enforced positivity using the absolute value of all network outputs such that $\hat{T}, \hat{Q}, \hat{K} \geq 0$. 

\begin{equation}
\boldsymbol{\overline{x}}
= \frac{\boldsymbol{x} - \boldsymbol{\mu}}{\boldsymbol{\sigma}}
\label{eq-normalize}
\end{equation}

In this physics-informed supervised regression task, we incorporated multiple loss terms. Temperature loss $L_T$, heat flow loss $L_Q$, and thermal conductivity loss $L_K$ were used to learn the measured data of thermal quantities, seen in Eqs. \ref{eq-LT}, \ref{eq-LQ}, and \ref{eq-LK}, where $V_{a}^{(train)}$ represents the training anchor nodes. Considering a finite and homogenous element (i.e., fixed average thermal conductivity $\overline{K}$) in three-dimensional space at depth $z$, we combined Eqs \ref{eq-conduction} and \ref{eq-heatflow} to quantify the deviation from the three-dimensional conductive heat transfer as a loss term $L_{pde}$, seen in Eq. \ref{eq-Lpde}. Furthermore, we noted that deep subsurface temperatures generally increase along depth given intervals of hundreds of meters (except for rare scenarios, such as subduction zones and deep aquifers undergoing convection to shallower depths \cite{ziagos1986model}), hence we introduced an additional physics-informed loss $L_{c}$, seen in Eq. \ref{eq-Lc}, to constrain our model accordingly. Note that derivatives were computed using automatic differentiation of neural networks \cite{baydin2018automatic}. Finally, the total loss $L$ was computed as the weighted sum using hyperparameters $\{\lambda_T, \lambda_Q, \lambda_K, \lambda_{pde}, \lambda_c\}$, seen in Eq. \ref{eq-l}.

\begin{equation}
L_T= \mathbb{E}_{v \sim V_{a}^{(train)}} \Big[ (\hat{T}_v - T_v)^2\Big]
\label{eq-LT}
\end{equation}

\begin{equation}
L_Q= \mathbb{E}_{v \sim V_{a}^{(train)}} \Big[ (\hat{Q}_v - Q_v)^2\Big]
\label{eq-LQ}
\end{equation}

\begin{equation}
L_K= \mathbb{E}_{v \sim V_{a}^{(train)}} \Big[ (\hat{K}_v - K_v)^2\Big]
\label{eq-LK}
\end{equation}

\begin{equation}
\begin{split}
& L_{pde} = \mathbb{E}_{v \sim V_g} \Bigg[\Bigg(\overline{K_v} \Big(\frac{\partial \hat{T}_v}{\partial x} + \frac{\partial \hat{T}_v}{\partial y} + \frac{\partial \hat{T}_v}
{\partial z}\Big) + \hat{Q}_v + \\
& \int_{0}^{z_v} \Bigg(\frac{\partial}{\partial x} \Big(\hat{K}_v \frac{\partial \hat{T}_v}{\partial x}\Big) + \frac{\partial}{\partial y} \Big(\hat{K}_v \frac{\partial \hat{T}_v}{\partial y}\Big) + \frac{\partial}{\partial z'} \Big(\hat{K}_v \frac{\partial \hat{T}_v}{\partial z'}\Big)\Bigg) dz'\Bigg)^2 \Bigg]\\
\end{split}
\label{eq-Lpde}
\end{equation}

\begin{equation}
L_c= \mathbb{E}_{v \sim V_g} \Big[ \max\Big(-\frac{\partial \hat{T}_v}{\partial z}, 0\Big)^2\Big]
\label{eq-Lc}
\end{equation}

\begin{equation}
L = \lambda_T L_T + \lambda_Q L_Q + \lambda_K L_K + \lambda_{pde} L_{pde} + \lambda_c L_c
\label{eq-l}
\end{equation}

Minimizing such multi-objective loss is an inherently challenging aspect of training physics-informed neural networks. In this work, we alleviated this challenge using a self-adaptive loss balancing scheme called Relative Loss Balancing with Random Lookback (ReLoBRaLo), seen in Eq. \ref{eq-loss-balancing} \cite{bischof2021multi}, where $\alpha$ is the exponential decay rate, $\rho$ is a Bernoulli random variable with $\mathbb{E}[\rho] \simeq 1$, $t$ is the current iteration, $t-1$ is the previous iteration, $m$ is the number of loss terms, and $\tau$ is temperature term. We used a uniform prior over loss weights at step zero and set $\alpha=0.9$, $\tau=0.1$, and $\mathbb{E}[\rho]=1$ as recommended in the original ReLoBRaLo paper \cite{bischof2021multi}.

\begin{equation}
\begin{split}
& \forall \ \ i \in \mathcal{L} = \{T, Q, K, pde, c\} \ and \ m = |\mathcal{L}|\\ 
& \lambda^{bal}_i(t,t') = m \cdot \frac{exp\Big(\frac{L_i(t)}{\tau L_i(t')}\Big)}{\sum_{j=1}^m exp\Big(\frac{L_j(t)}{\tau L_j(t')})} \\
& \lambda^{hist}_i(t) = \rho \lambda_i(t-1) + (1-\rho)\lambda^{bal}_i(t,0) \\
& \lambda_i(t) = \alpha \lambda^{hist}_i(t) + (1-\alpha) \lambda^{bal}_i(t,t-1)
\end{split}
\label{eq-loss-balancing}
\end{equation}

In this study, we minimized $L$ using the Adam optimizer \cite{kingma2014adam} for a total of $400$ epochs, with initial learning rate of $0.001$, which was reduced when the validation loss stopped to improve. Unlike uncorrelated data structures, records (i.e., nodes) are interdependent in graph settings where the prediction for a given node is affected by its neighbourhood. In sampling a batch of nodes, we created a node-induced subgraph. However, especially in large and sparse graphs, it is very likely that the neighbourhood of the sampled nodes would be misrepresented in this subgraph. Consequently, the InterPIGNN model would not be able to properly leverage neighbourhood information during training, which would hinder the learning process. Our graph is fairly large with $|V| = n = 4,679,670$ nodes and $|E|=23,398,350$ edges, so it is extremely sparse (i.e., graph density is defined as $|E|/(|V|(|V|-1))$). To resolve this challenge, we further included the neighbourhood of those already randomly sampled batch nodes.

\subsection{Uncertainty Quantification}\label{sub3.4}
Quantifying uncertainty in subsurface engineering is crucial for decision making and risk minimization, budgeting and investment allocation, resource management, modeling and simulation, regulatory compliance, amongst other considerations. Generally, there are two types of uncertainties: aleatoric or data uncertainty, and epistemic or model uncertainty \cite{gal2016uncertainty, hullermeier2021aleatoric, kendall2017uncertainties}. In this work, we focused on estimating the epistemic uncertainty associated with the trained model using the Monte Carlo Dropout \cite{gal2016dropout}. This approach provides a theoretical framework showing that an arbitrary neural network with dropout applied before every weight layer, is mathematically equivalent to an approximation to the probabilistic deep Gaussian process \cite{damianou2013deep}. 

We can directly apply Monte Carlo Dropout using our trained model with no changes since dropout layers are already included prior to each weight layer. To estimate the predictive mean and predictive uncertainty we simply collected the results of stochastic forward passes through the model. Note that these forward passes are done concurrently, resulting in constant running time identical to that of standard dropout. We note that the size of the variance of the posterior distribution generated by Monte Carlo Dropout mainly depends on (1) dropout rate, (2) the size and architecture of the neural network, and (3) magnitude of the output quantity. Nevertheless, the posterior variance is independent of the input data size and variance \cite{verdoja2020notes}.

\subsection{Prediction Explainability}\label{sub3.5}
Beyond achieving accurate predictions of subsurface temperature, we were also interested in understanding which and how input quantities (i.e., features) are most important to model predictions. Model explainability can be local or global, where the former motivate the prediction of a given instance while the latter provides insights into model predictions across the entire input domain. The machine learning literature shows a variety of explainability techniques, such as Gradients, Grad-CAM, Guided BackPropagation, DeepLIFT, Deconvolution, amongst others \cite{linardatos2020explainable, simonyan2013deep, sundararajan2017axiomatic, shrikumar2017learning, springenberg2014striving, zeiler2014european, selvaraju2017grad}. Other approaches provide axiomatic attribution that goes beyond using feature values and gradients at the input vector. Integrated Gradients is an attribution algorithm developed on axiomatic basis \cite{sundararajan2017axiomatic}. Particularly, this algorithm satisfies six axioms:

\begin{enumerate}
    \item \textit{Completeness}: given an input vector, the sum of feature attributions is equal to the difference between the model outputs for that input vector and a baseline vector.
    \item \textit{Implementation Invariance}: if two neural networks compute identical functions for all inputs, they receive identical attributions.
    \item \textit{Sensitivity}: if an input vector differs from a baseline vector in a single variable resulting in a distinct output then that variable should be attributed accordingly
    \item \textit{Insensitivity}: no attribution is assigned to a variable that has no impact on the output..
    \item \textit{Linearity Preservation}: the attribution of a linear combination of input vectors is equal to the linear combination of the attributions of the individual input vectors.
    \item \textit{Symmetry}: symmetric variables with identical values receive equal attributions.
\end{enumerate}

Theoretically, the Integrated Gradients method represents the unique path integral from a baseline vector to an input vector of interest satisfying the aforementioned axioms. Historically, the Integrated Gradients algorithm is the Aumann-Shapley method from cooperative game theory, which holds similar characteristics \cite{friedman2004paths}. We apply Integrated Gradients on the model prediction of each of the $4,279,536$ grid nodes to capture local feature attributions. The magnitudes of these attributions are then aggregated to understand the cumulatively most influential features.

\section{Results}\label{sec4}
\begin{figure}[h!]%
\centering
\includegraphics[width=\textwidth]{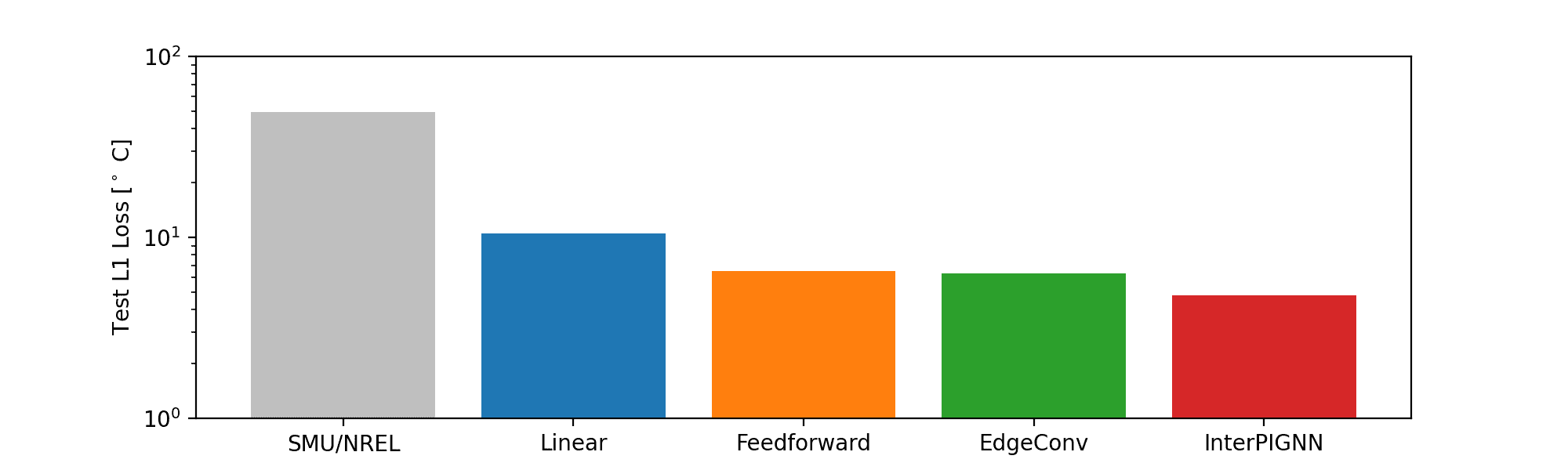}
\caption{Comparison of the mean absolute error of the predicted temperature-at-depth on the $10\%$ hold-out test set for different algorithms (smaller values are better).}\label{fig_model_errors}
\end{figure}

We first examined the performance of our InterPIGNN model compared to the combined SMU/NREL temperature-at-depth model, linear regression, feedforward neural network, and EdgeConv point-based GNN. Except for the combined SMU/NREL model, all of these models make use of the physical quantities described in this study. Whereas these algorithms have inherently different architectures, we ensured that they were trained and tested using the same preprocessing and postprocessing steps. Seen in Fig. \ref{fig_model_errors}, we compared the different models using the mean absolute error based on the $10\%$ hold-out test set and found that the combined SMU/NREL mode, linear regression, feedforward, EdgeConv, and InterPIGNN achieved $49.5$, $13.0$, $6.5$, $6.3$, and $4.8^\circ C$, respectively. We first note that our models are superior to the combined SMU/NREL model, which indicates that the considered physical quantities are overall useful predictors of BHT. Whereas using GNN algorithms was advantageous, the proposed InterPIGNN model was found to be the most accurate. Comparing the InterPIGNN and EdgeConv architectures, we can theoretically attribute this improved InterPIGNN performance to the inclusion of the target quantities (e.g., temperature, surface heat flow, and rock thermal conductivity) as neighbouring features in the message-passing computation. This was further confirmed empirically through feature explainability, as described later. Additionally, InterPIGNN simultaneously generates predictions of surface heat flow and rock thermal conductivity which were found to have mean absolute errors of $5.817 \ mW/m^2$ and $0.022 \ W/(C \cdot m)$, respectively. Meanwhile, our model closely satisfied the the three-dimensional conductive heat transfer loss with mean absolute error of $0.9 \ mW/m^2$, and resulted in non-decreasing temperatures with depth almost everywhere.

\begin{figure}[h!]%
\centering
\includegraphics[width=\textwidth]{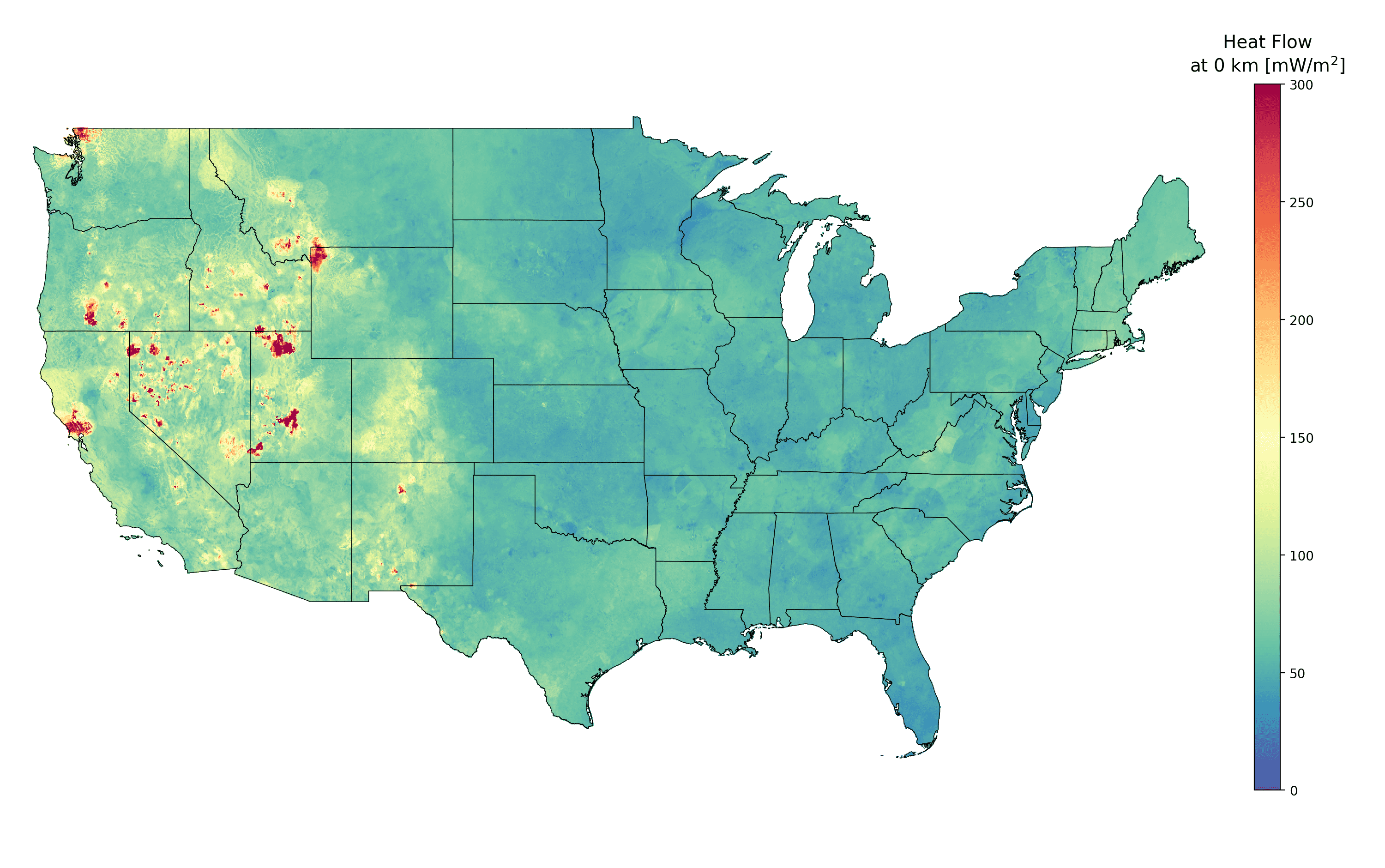}
\caption{Predicted surface heat flow map.}\label{fig_Heat_Flow_0km}
\end{figure}

Fig. \ref{fig_Heat_Flow_0km} shows the predicted surface heat flow across the continental United States. Heat flow values were limited to a range of $0-300 \ mW/m^2$ for visual convenience. Whereas the SMU maps conveniently excluded heat flow measurements greater than $120 \ mW/m^2$, our model overcomes this limitation by incorporating all heat flow measurements. The minimum, mean, and maximum predicted surface heat flow values were $17 \ mW/m^2$, $68 \ mW/m^2$, and $3,754 \ mW/m^2$, respectively. Seen in Fig. \ref{fig_Heat_Flow_0km}, our predictions show elevated surface heat flow across the Western United States largely due to the presence of the Pacific Ring of Fire where tectonic movements favor volcanic activities and geothermal heat transfer. Particularly, multiple locations stand out such as Yellowstone in Wyoming, Basin and Range Province extending mainly across Nevada and neighbouring states with thinner crustal thickness, Cascade Range starting from northern California and reaching to British Columbia, Canada and including active volcanoes like Mount St. Helens and Mount Rainier, Coso Volcanic Field in California, Rio Grande Rift stretching from New Mexico to Colorado, and Imperial Valley in California partly due to the San Andreas Fault system. Away from the Western flank, elevated surface heat flow values were also observed at the Appalachian Mountains and along the Gulf Coast.

% Figs. \ref{fig_Thermal_Conductivity_1km}, \ref{fig_Thermal_Conductivity_2km}, \ref{fig_Thermal_Conductivity_3km}, \ref{fig_Thermal_Conductivity_4km}, \ref{fig_Thermal_Conductivity_5km}, \ref{fig_Thermal_Conductivity_6km}, and \ref{fig_Thermal_Conductivity_7km} show the spatial spread of thermal conductivity predictions. For visual convenience, values were bounded to a range of $1.0-5.0 \ W/(C \cdot m)$. 

\begin{figure}[h!]%
\centering
\includegraphics[scale=0.65]{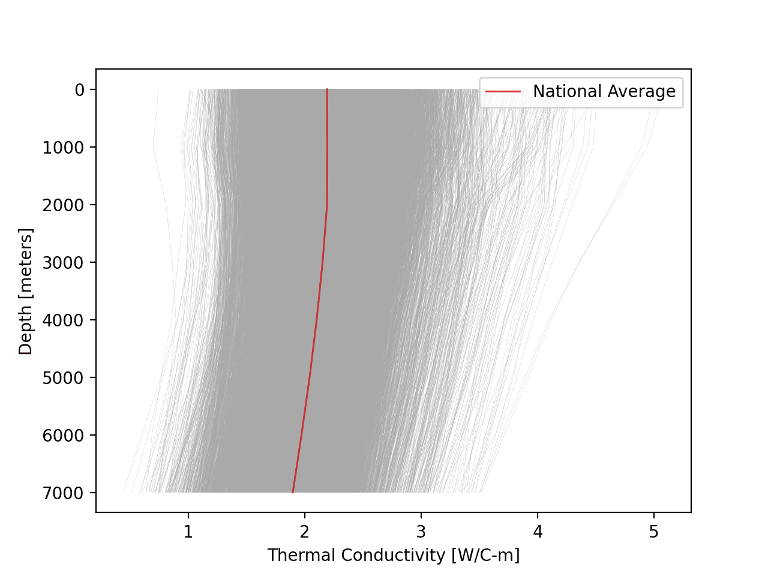}
\caption{Subsurface thermal conductivity profiles across grid nodes $V_g$ (gray lines) and on average (red line).}\label{fig_thermal_conductivity_profile}
\end{figure}

The minimum, mean, and maximum predicted rock thermal conductivity values were $0.099 \ W/(C \cdot m)$, $2.093 \ W/(C \cdot m)$, and $5.512 \ W/(C \cdot m)$, respectively. Unlike the diffusive heat flow and temperature quantities, thermal conductivity is a rock property with compartmentalized distribution as can be observed in the raw measurements and model predictions. It is governed by various parameters such as rock composition, porosity, fluid saturation, temperature, and pressure. It tends to be lower in dry and unconsolidated rocks and varies with rock composition \cite{song2023influencing}. Seen in Fig. \ref{fig_thermal_conductivity_profile}, our predictions show a clear trend of decreasing rock thermal conductivity along depth. This is attributed to the increase in temperature with depth which, in most rock types, result in decreasing rock thermal conductivity \cite{chen2021effect}.

\begin{figure}[h!]%
\centering
\includegraphics[width=\textwidth]{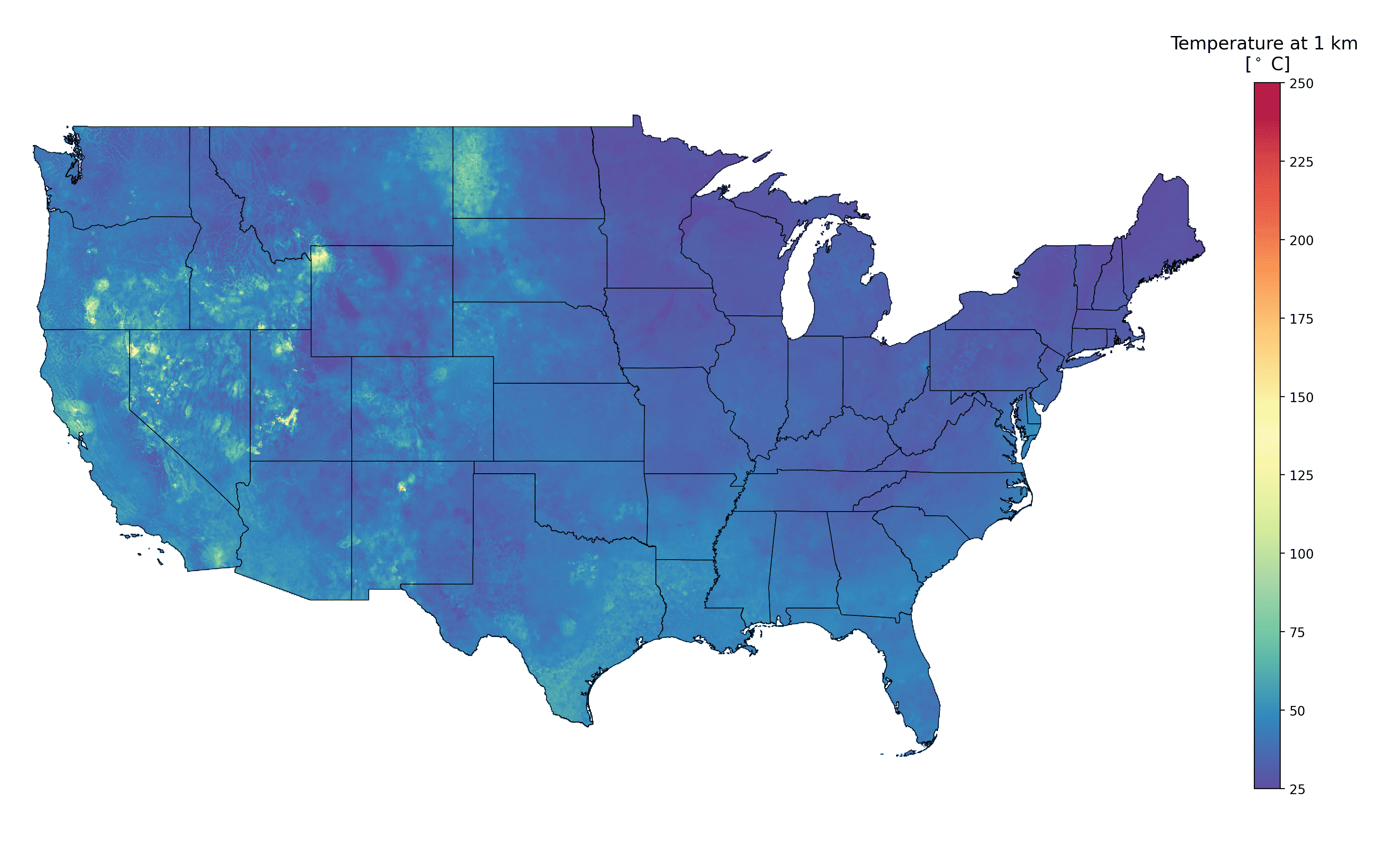}
\caption{Predicted temperature-at-depth map at a depth of 1 km.}\label{fig_1km}
\end{figure}

\begin{figure}[h!]%
\centering
\includegraphics[width=\textwidth]{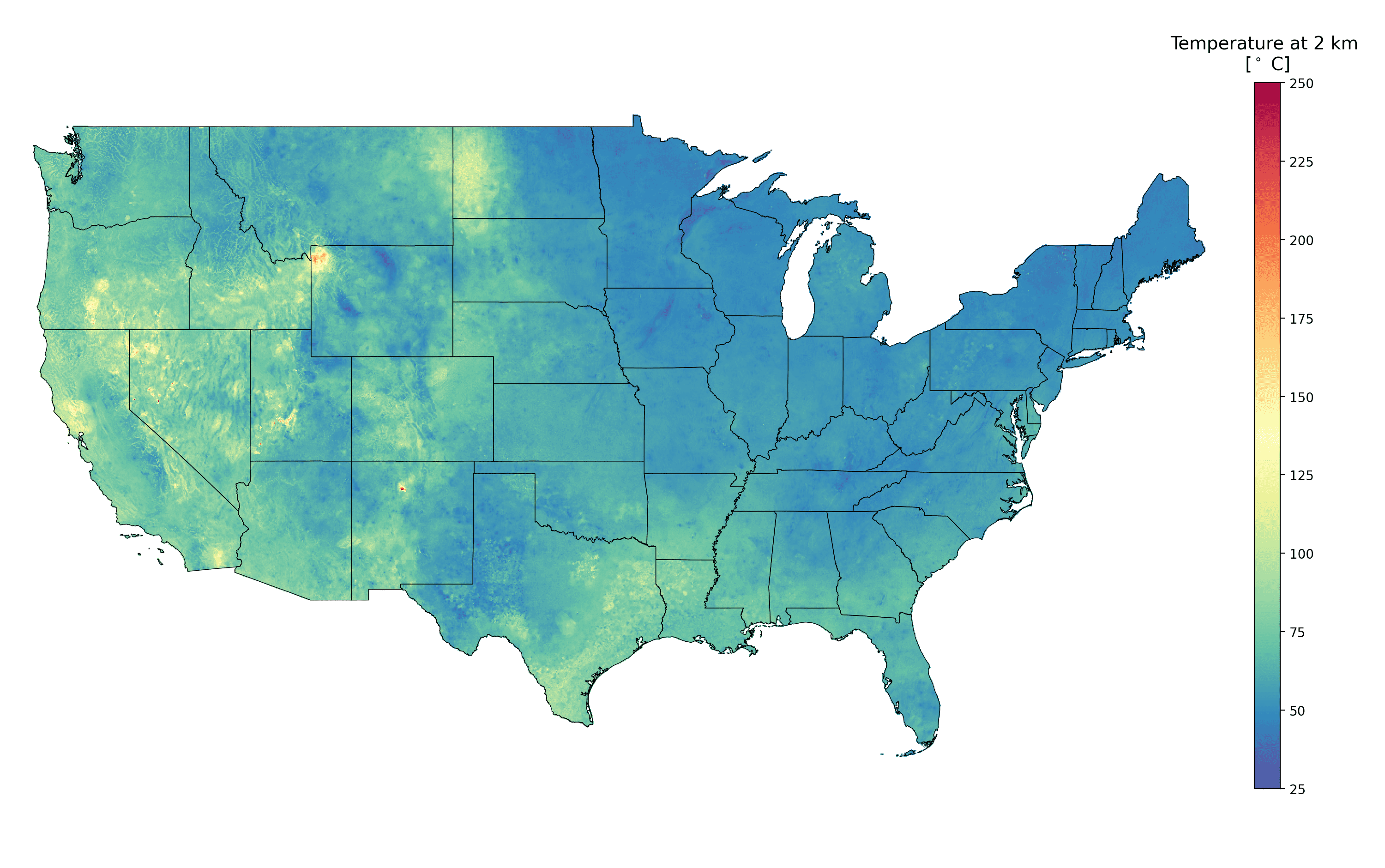}
\caption{Predicted temperature-at-depth map at a depth of 2 km.}\label{fig_2km}
\end{figure}

\begin{figure}[h!]%
\centering
\includegraphics[width=\textwidth]{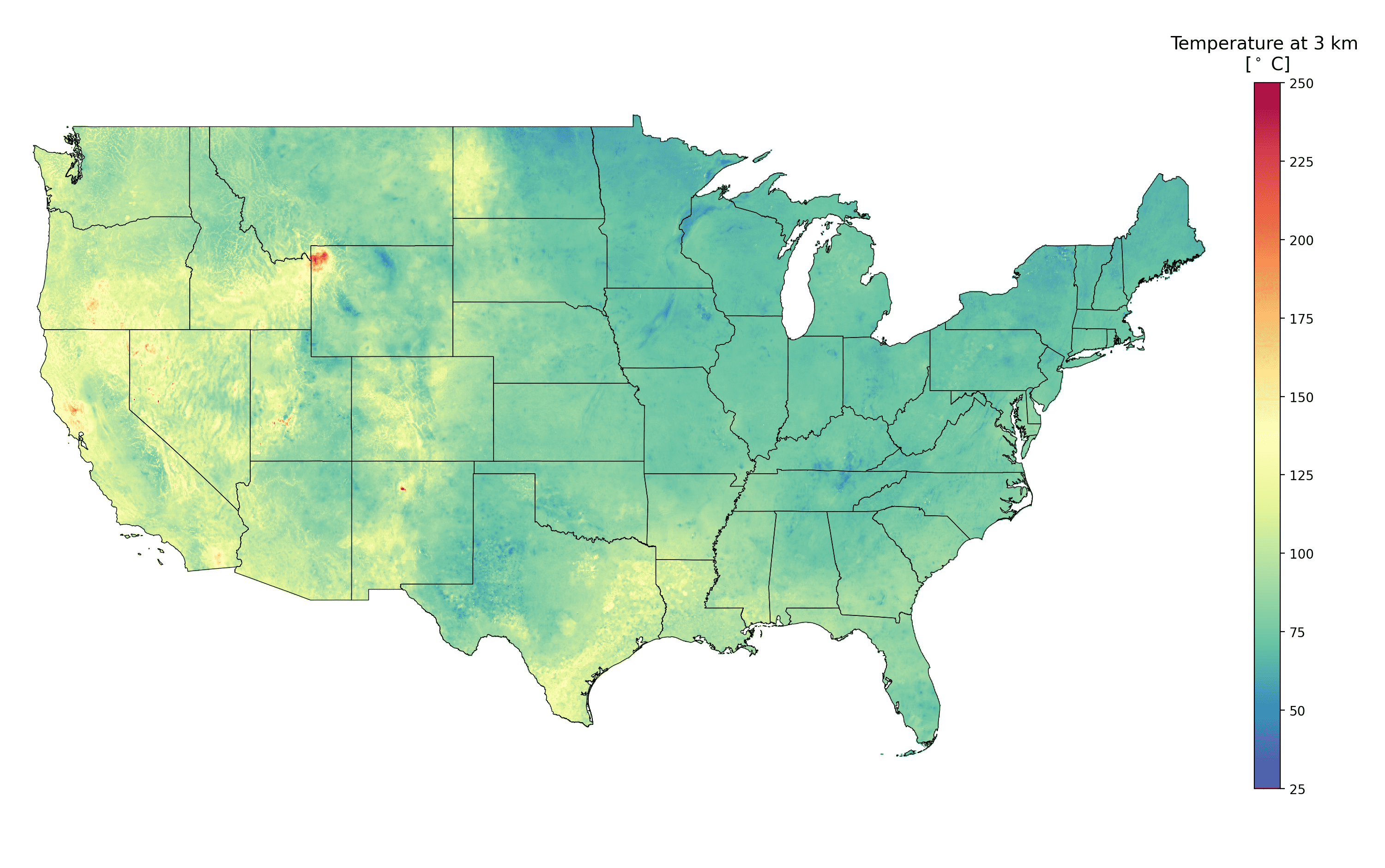}
\caption{Predicted temperature-at-depth map at a depth of 3 km.}\label{fig_3km}
\end{figure}

\begin{figure}[h!]%
\centering
\includegraphics[width=\textwidth]{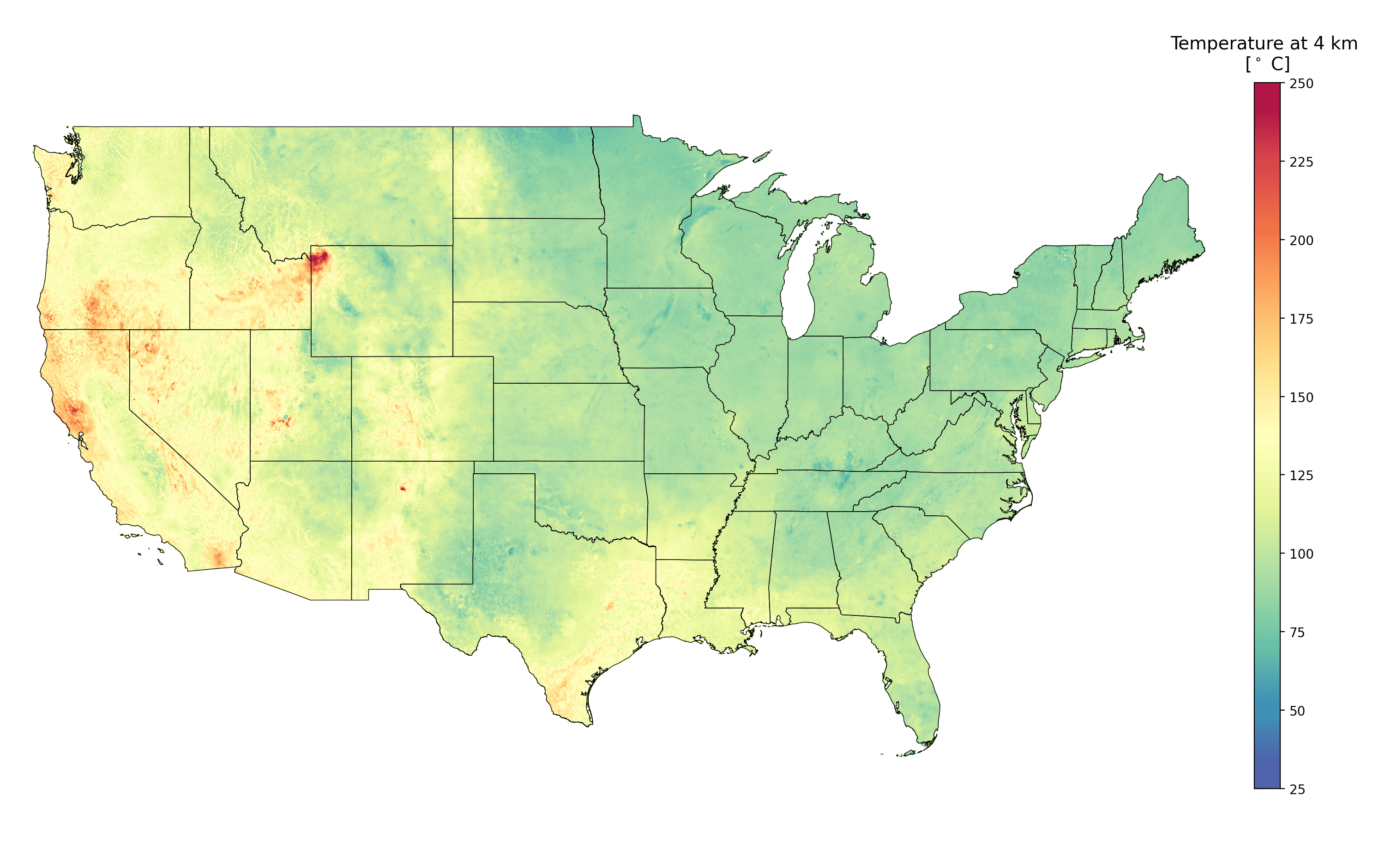}
\caption{Predicted temperature-at-depth map at a depth of 4 km.}\label{fig_4km}
\end{figure}

\begin{figure}[h!]%
\centering
\includegraphics[width=\textwidth]{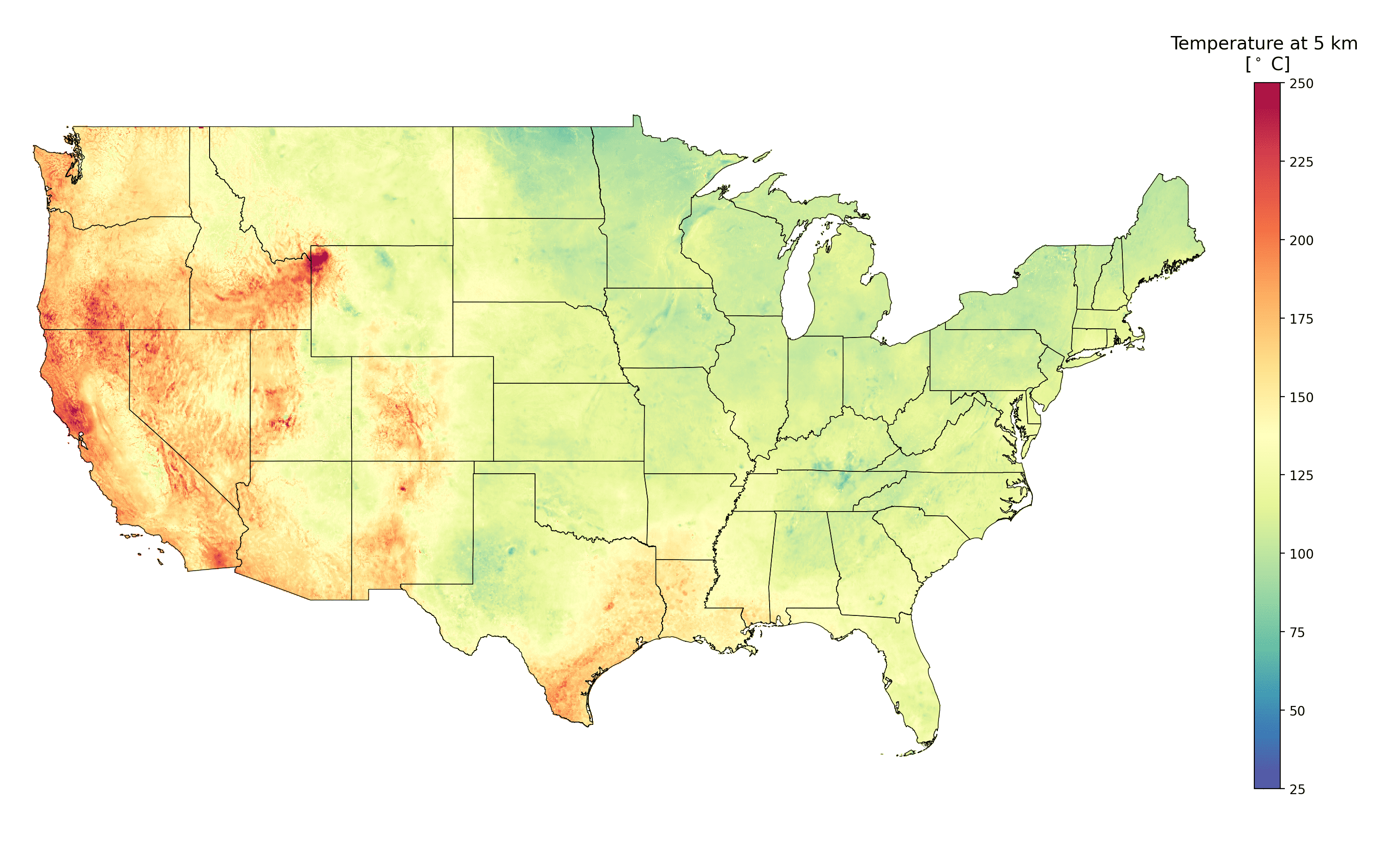}
\caption{Predicted temperature-at-depth map at a depth of 5 km.}\label{fig_5km}
\end{figure}

\begin{figure}[h!]%
\centering
\includegraphics[width=\textwidth]{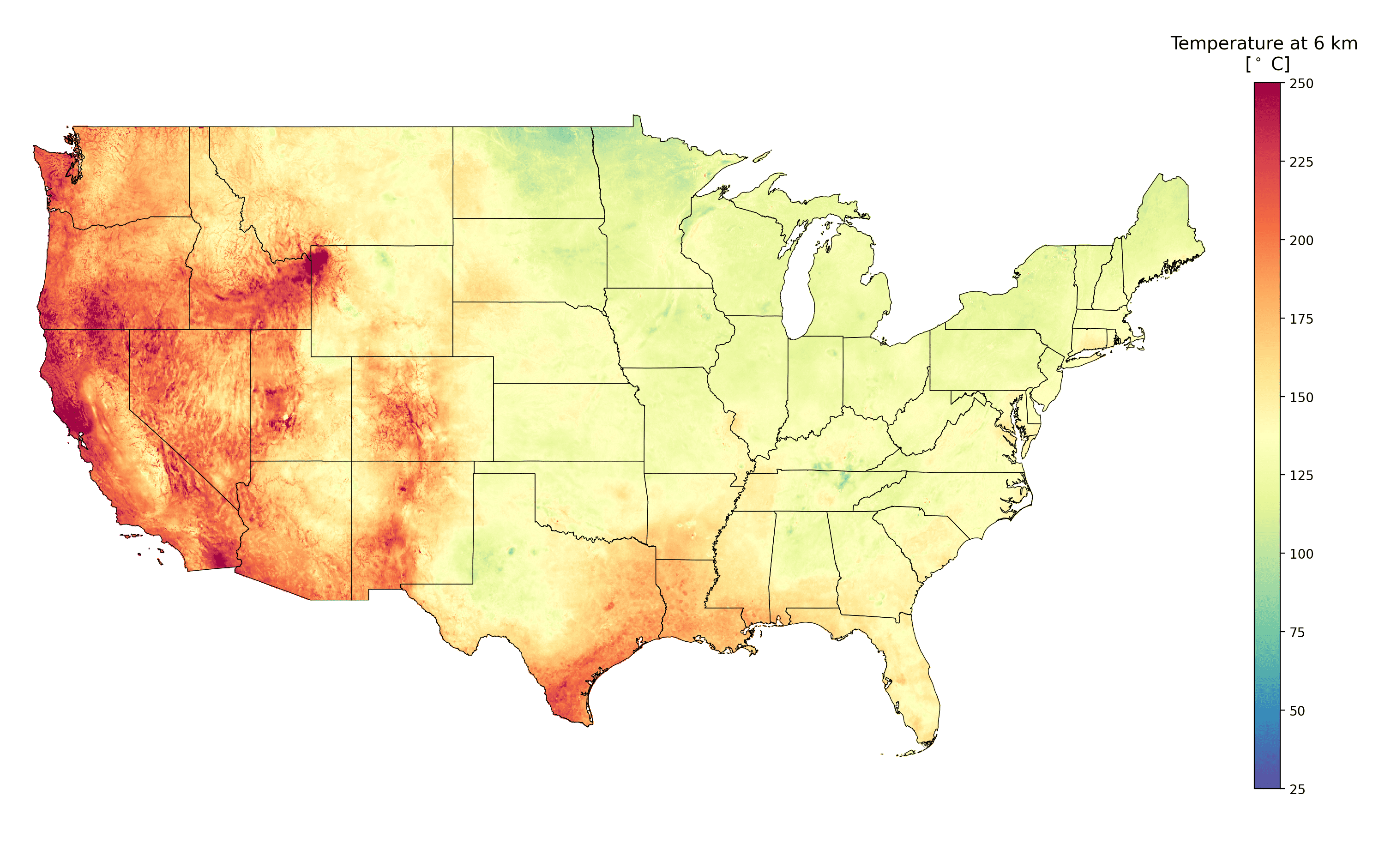}
\caption{Predicted temperature-at-depth map at a depth of 6 km.}\label{fig_6km}

\end{figure}

\begin{figure}[h!]%
\centering
\includegraphics[width=1.0\textwidth]{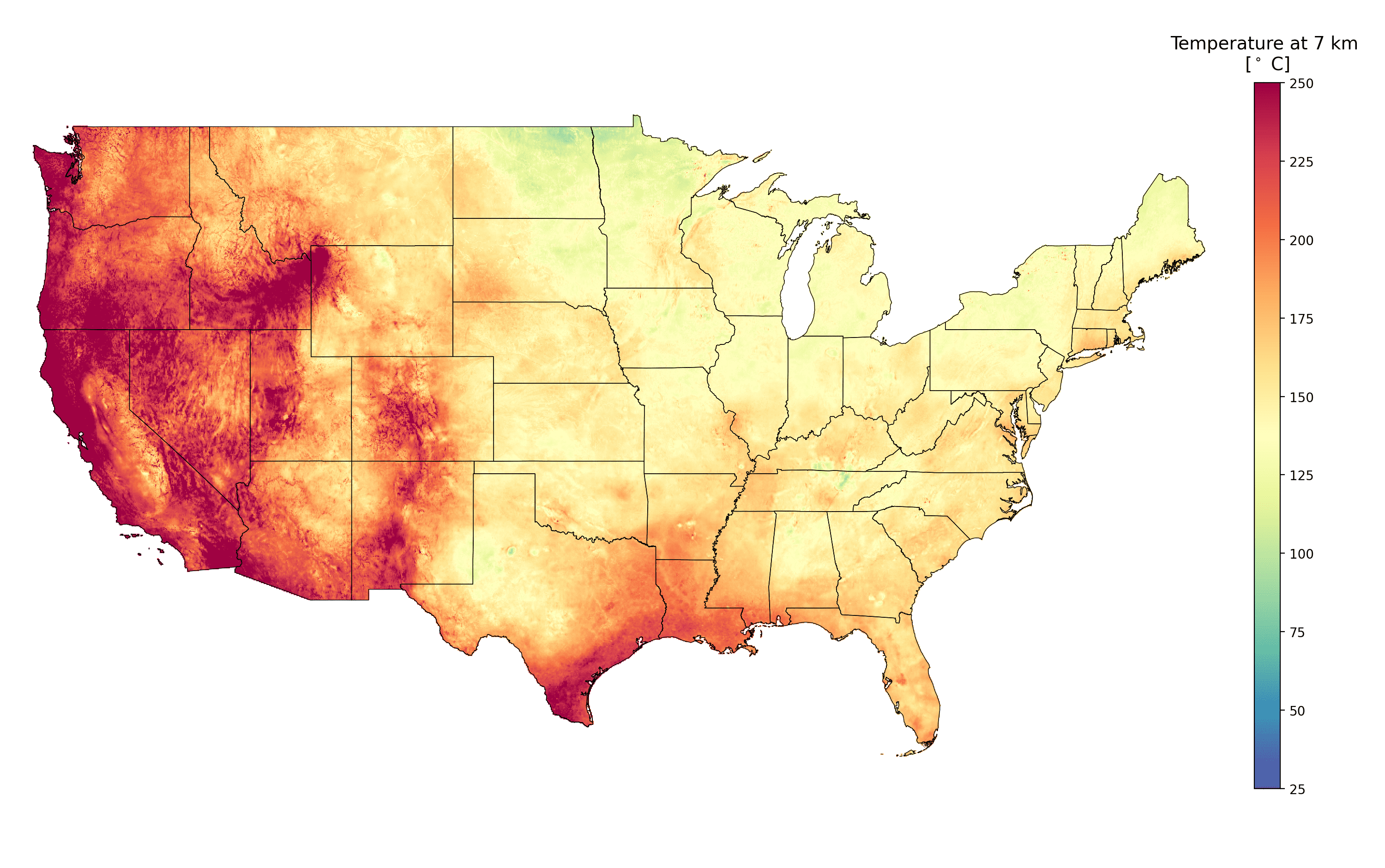}
\caption{Predicted temperature-at-depth map at a depth of 7 km.}\label{fig_7km}
\end{figure}

Figs. \ref{fig_1km}, \ref{fig_2km}, \ref{fig_3km}, \ref{fig_4km}, \ref{fig_5km}, \ref{fig_6km}, and \ref{fig_7km} show the predicted temperature-at-depth across the conterminous United States for depths of 1-7 km, respectively. Temperatures were limited to a range of $25-250^\circ C$ for visual convenience. The greatest temperature predicted at a depth of 7 km was at the Yellowstone Caldera with a magnitude of $381^\circ C$. As anticipated, other high-temperature spots were also observed around volcanic and thermally active areas, such as Valles Caldera in New Mexico, La Garita Caldera in Colorado, Newberry Volcano in Oregon, The Great Basin, The Geysers in California, Coso Volcanic Field in California, Salton Buttes in California, amongst others.

\begin{figure}[h!]%
\centering
\includegraphics[scale=0.65]{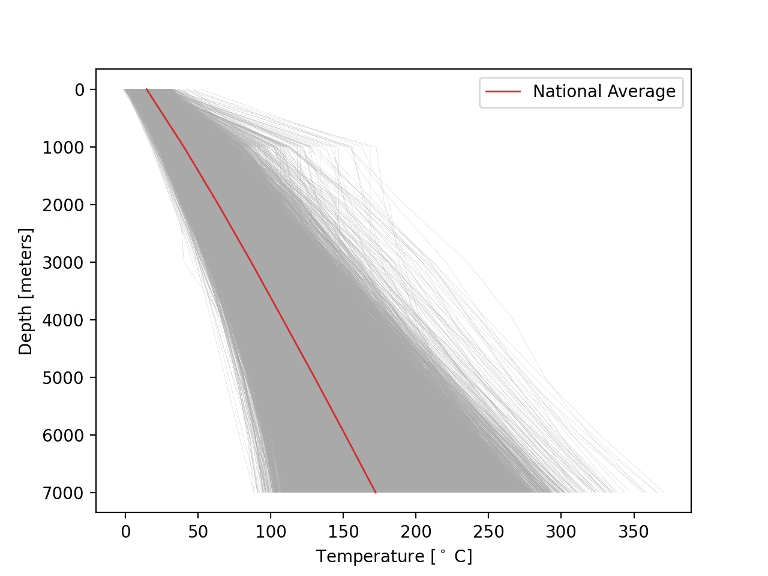}
\caption{Subsurface temperature profiles across grid nodes $V_g$ (gray lines) and on average (red line).}\label{fig_temperature_profile}
\end{figure}

\begin{figure}[h!]%
\centering
\includegraphics[scale=0.65]{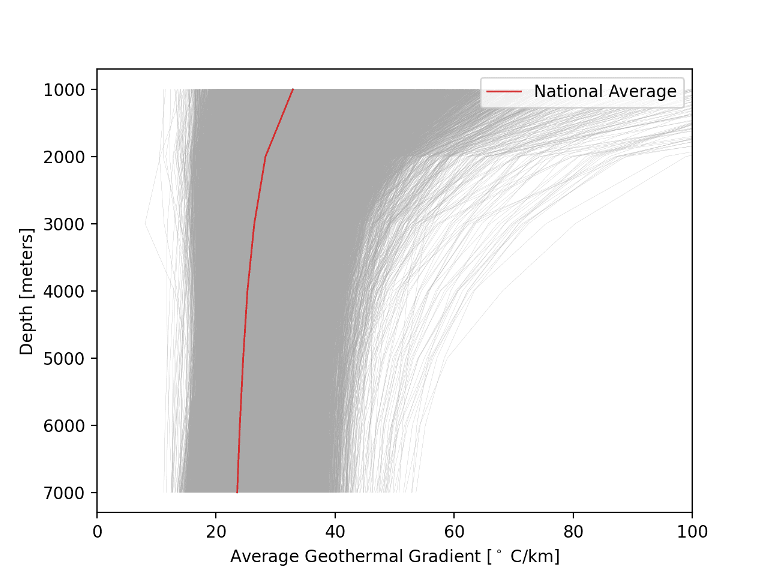}
\caption{Average geothermal gradient profiles across grid nodes $V_g$ (gray lines) and on average (red line).}\label{fig_geothermal_gradient_profile}
\end{figure}

\begin{figure}[h!]%
\centering
\includegraphics[scale=0.65]{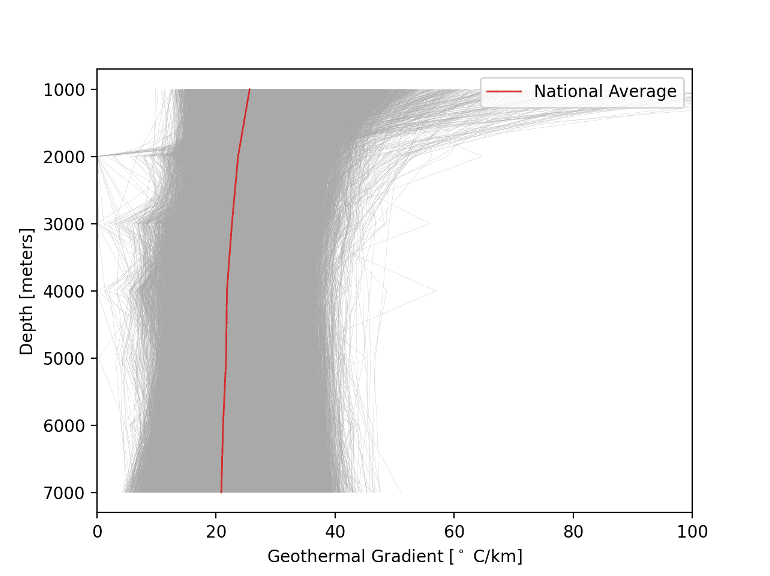}
\caption{Local geothermal gradient profiles across grid nodes $V_g$ (gray lines) and on average (red line).}\label{fig_local_geothermal_gradient_profile}
\end{figure}

As seen in Eqs. \ref{eq-Lpde}-\ref{eq-Lc}, the InterPIGNN model was constrained to respect three-dimensional conductive heat transfer and the fact that temperature generally increases with depth. This resulted in physical predictions as seen in Fig. \ref{fig_temperature_profile} which shows the predicted subsurface temperature profiles across various locations (gray lines) as well as the average profiles (red line). We observe the desired behavior of strictly increasing temperatures with depth with declining geothermal gradients. Seen in Figs. \ref{fig_geothermal_gradient_profile}-\ref{fig_local_geothermal_gradient_profile}, the average and local geothermal gradient profiles can be visualized more closely across locations (gray lines) as well as on average (red line). We generally observe decreasing average geothermal gradients with depth, which is aligned with the established finding that radiogenic heat generation in the Earth's crust exponentially declines with depth \cite{brady2006distribution}.

\begin{figure}[h!]%
\centering
\includegraphics[scale=0.65]{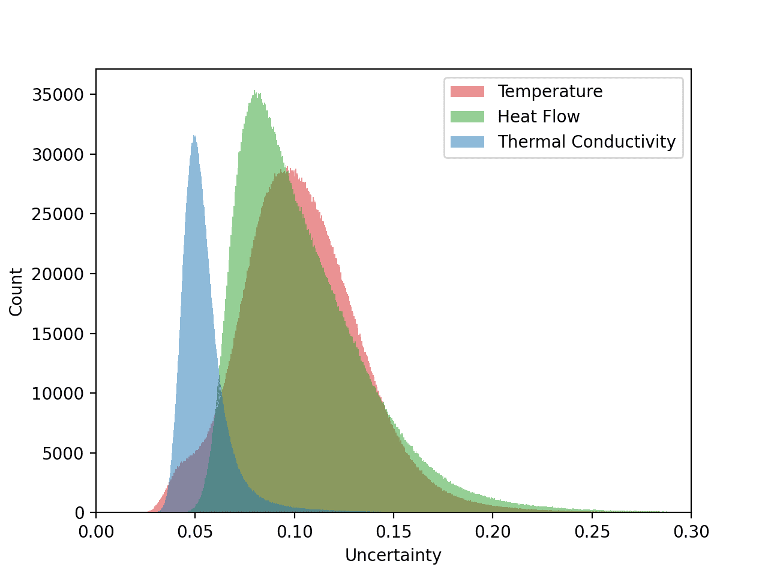}
\caption{Temperature-at-depth uncertainty distribution as modeled using Monte Carlo Dropout.}\label{fig_uncertainty}
\end{figure}

With the uncertainties surrounding subsurface physical quantities, we used Monte Carlo Dropout to estimate the model epistemic uncertainty associated with our predictions of temperature, surface heat flow, and rock thermal conductivity. Fig \ref{fig_uncertainty} shows the distribution of uncertainty across our target thermal quantities. The median uncertainty in temperature, surface heat flow, and rock thermal conductivity was $10.3\%$, $9.7\%$, and $5.2\%$, respectively.

\begin{figure}[h!]%
\centering
\includegraphics[scale=0.25]{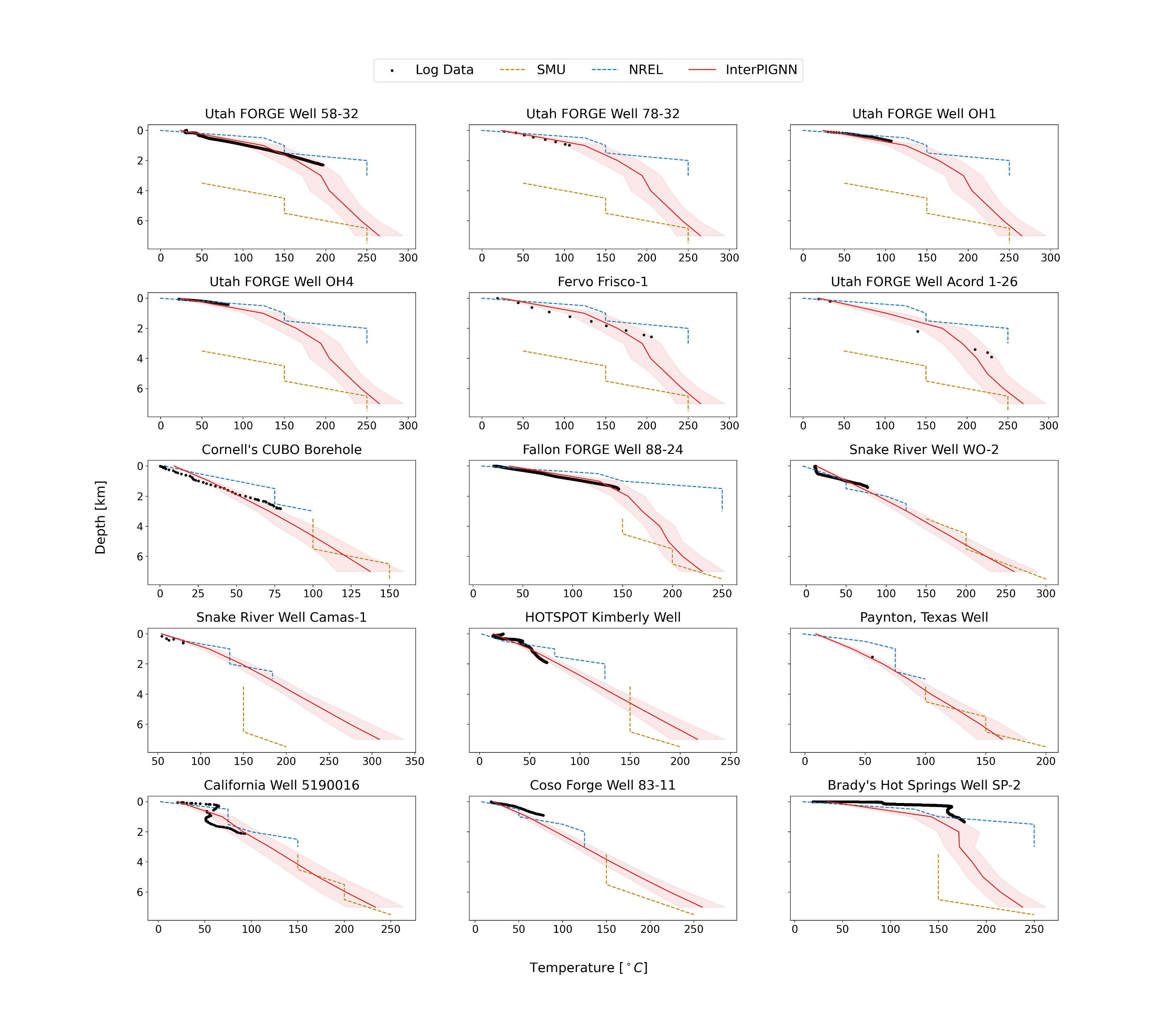}
\caption{Verification of the proposed InterPIGNN model using temperature log data across different locations and projects. Whereas the solid line represents the InterPIGNN model prediction, the shaded area represents the uncertainty band. the NREL and SMU models were also visualized in dashed lines.}\label{fig_temperature_log_comparison}
\end{figure}

To verify the model performance and demonstrate its utility, we evaluated its predictions against measured temperature log data. This involved 15 wells from multiple geothermal projects: Utah FORGE wells 58-32, 78-32, OH1, OH4, and Acord 1-26, Fervo Energy well Frisco-1, Cornell University Borehole Observatory (CUBO) well, Fallon Forge 88-24, Snake River Wells WO-2 and Camas-1, HOTSPOT Kimberly well, Paynton, Texas well, California well 5190016, Coso Forge well 83-11, and Brady's Hot Springs well SP-2 \cite{allis2018thermal, cladouhos2015newberry, purwamaskapreliminary, podgorney1991snake, blankenship2017proposed, shervais2013first}. In the case of temperature logs which were run soon after well drilling and completion, we applied the Harrison correction \cite{harrison1983geothermal} to account for the disequilibrated thermal conditions in the wellbore as a result of drilling and completion fluid circulation. It is important to note that these corrections are correlations derived based on data measured in the field before and after thermal equilibrium across wellbores. Thus, the log temperature data might be a few degrees Celsius off compared to the actual rock temperatures at those locations. We also included the NREL and SMU predictions to demonstrate how they differ from our model. Seen in Fig. \ref{fig_temperature_log_comparison}, the InterPIGNN model predictions closely matched most of the temperature log data. Due to the increasing spatial sparsity of BHT measurements along depth, our model showed reasonable uncertainty magnitudes that were generally increasing with depth. We note that our model was superior to the NREL model, where the latter often overestimates subsurface temperatures. Additionally, unlike the SMU model whose predictions tend to frequently show static temperatures along deep intervals, our model shows increasing temperatures along depth.

\FloatBarrier
\section{Conclusion}\label{sec5}
We aggregated and processed BHT measurements for the conterminous US from multiple sources. To construct a thermal Earth model using these records, we introduced a novel interpolative physics-informed graph neural network module, named InterPIGNN, that is suitable for the interpolation of point cloud data structures.  We constructed surface heat flow, and temperature and thermal conductivity predictions for depths of $0$-$7 \ km$ at an interval of $1 \ km$ with spatial resolution of $18 \ km^2$ per grid cell. Our model showed superior temperature, surface heat flow and thermal conductivity mean absolute errors of $4.8^\circ C$, $5.817 \ mW/m^2$ and $0.022 \ W/(C \cdot m)$, respectively. We also found that elevation, crustal thickness, depth, electric conductivity and seismic features were most important to the model predictions. We additionally visualized local feature attributions to provide further understanding of the model predictions across three-dimensional space.

\FloatBarrier

\backmatter

\bmhead{Data Availability}
It is also available as feature layers on ArcGIS at \href{https://arcg.is/nLzzT0}{https://arcg.is/nLzzT0}. 

\bmhead{Conflict of Interest}
The authors have no conflicts of interest to declare that are relevant to the content of this article.

\bmhead{Ethical Statement}
This material is the authors' own original work, which has not been previously published elsewhere. This paper is not currently being considered for publication elsewhere.  The paper reflects the authors' own research and analysis in a truthful and complete manner. The paper properly credits the meaningful contributions of co-authors and co-researchers. The results are appropriately placed in the context of prior and existing research. All sources used are properly disclosed. All authors have been personally and actively involved in substantial work leading to the paper, and will take public responsibility for its content.

\bmhead{Funding Declaration}
No funding was received to assist with the preparation of this manuscript.

\bmhead{Author Contribution}
\textbf{Mohammad Aljubran} contributed to conceptualization, methodology, implementation and analysis, and writing first draft. \textbf{Roland Horne} contributed to conceptualization, research supervision, and reviewing and editing.

\clearpage

% \bibliography{sn-bibliography}% common bib file

\end{document}